\documentclass[10pt,twocolumn]{article}

\usepackage[a4paper,margin=0.75in]{geometry}
\usepackage[T1]{fontenc}
\usepackage{microtype}
\usepackage[hidelinks]{hyperref}
\usepackage{url}
\usepackage{graphicx}
\usepackage{amsmath,amssymb}
\usepackage{booktabs}
\usepackage{siunitx}
\usepackage{tikz}
\usepackage{todonotes}
\usepackage{acro}

\acsetup{   
  list/template = tabular,
  list/display = used,
  list/heading = none,
  format/short=\normalfont
}

\usepackage{titlesec}
\titlespacing*{\section}{0pt}{8pt}{1pt}
\titlespacing*{\subsection}{0pt}{6pt}{1pt}
\titlespacing*{\subsubsection}{0pt}{5pt}{1pt}

\usepackage[font=small,labelfont=bf,labelsep=period]{caption}
\usepackage{subcaption}

\titleformat{\paragraph}
  {\normalsize\bfseries}{\theparagraph}{0pt}{}
\titlespacing*{\paragraph}{0pt}{6pt}{0pt}

\usepackage{fancyhdr}
\usepackage{authblk}

\makeatletter
\renewcommand\AB@affilsepx{;\space}
\makeatother

\makeatletter

\renewcommand{\footnoterule}{%
  \kern-3pt
  \hrule width \columnwidth height 0.3pt
  \kern 2.6pt
}
\newcommand{\titlehrule}{%
  \vspace{0.6em}
  \hrule height 0.4pt
  \vspace{0.6em}
}

\newcommand{\correspondence}[1]{\gdef\@correspondence{#1}}
\correspondence{}

\newcommand{\abstracttext}[1]{\gdef\@abstract{#1}}
\abstracttext{}

\AtBeginDocument{%
\renewcommand{\maketitle}{%
  \vspace*{0pt}
  \noindent
  \begin{minipage}{\textwidth}

    {\bfseries\LARGE \@title\par}
    \vspace{1cm}

    \titlehrule

    {\normalsize \AB@authlist\par}
    \vspace{0.2cm}

    {\footnotesize
    \makeatletter
    \AB@affillist%
    \makeatother
    \ifx\@correspondence\@empty\relax\else
      ;\space \@correspondence
    \fi
    \par}

    \vspace{0.5cm}

    \titlehrule

    \small
    \@abstract
    \vspace{0.7cm}

  \end{minipage}
}
}
\makeatother

\DeclareAcronym{gbm}{
  short = GBM ,
  long  = glioblastoma
}

\DeclareAcronym{os}{
  short = OS ,
  long  = overall survival
}

\DeclareAcronym{rt}{
  short = RT ,
  long  = radiotherapy
}

\DeclareAcronym{soc}{
  short = SOC ,
  long  = standard-of-care
}

\DeclareAcronym{ctv}{
  short = CTV ,
  long  = clinical target volume
}

\DeclareAcronym{tcia}{
  short = TCIA ,
  long  = The Cancer Imaging Archive
}

\DeclareAcronym{brats}{
  short = BraTS ,
  long  = Brain Tumor Segmentation
}

\DeclareAcronym{com}{
  short = CoM ,
  long  = center of mass
}

\DeclareAcronym{adc}{
  short = ADC ,
  long  = apparent diffusion coefficient
}

\DeclareAcronym{dti}{
  short = DTI ,
  long  = diffusion tensor imaging
}

\DeclareAcronym{syn}{
  short = SyN ,
  long  = symmetric normalization
}

\DeclareAcronym{t1c}{
  short = T1c ,
  long  = contrast-enhanced T1-weighted
}

\DeclareAcronym{t1}{
  short = T1 ,
  long  = T1-weighted
}

\DeclareAcronym{t2}{
  short = T2 ,
  long  = T2-weighted
}

\DeclareAcronym{flair}{
  short = FLAIR ,
  long  = fluid-attenuated inversion recovery
}

\DeclareAcronym{csf}{
  short = CSF ,
  long  = cerebrospinal fluid
}

\DeclareAcronym{odil}{
  short = ODIL ,
  long  = Optimizing a Discrete Loss
}

\DeclareAcronym{lmi}{
  short = LMI ,
  long  = Learn-Morph-Infer
}

\DeclareAcronym{pinn}{
  short = PINN ,
  long  = physics-informed neural network
}

\DeclareAcronym{loti}{
  short = LOTI ,
  long  = Lightweight Optimization for Estimating Tumor Infiltration
}

\DeclareAcronym{bce}{
  short = BCE ,
  long  = binary cross-entropy
}

\DeclareAcronym{pcc}{
  short = PCC ,
  long  = progression coverage coefficient
}

\DeclareAcronym{md}{
  short = MD ,
  long  = mean diffusivity
}

\DeclareAcronym{cptac}{
  short = CPTAC-GBM,
  long = Clinical Proteomic Tumor Analysis Consortium Glioblastoma Multiforme Collection
}

\DeclareAcronym{ivygap}{
  short = IVYGAP,
  long = Ivy Glioblastoma Atlas Project
}

\DeclareAcronym{tcgagbm}{
  short = TCGA-GBM,
  long = The Cancer Genome Atlas Glioblastoma Multiforme
}

\DeclareAcronym{upenngbm}{
  short = UPENN-GBM,
  long = Multi-parametric magnetic resonance imaging scans for de novo Glioblastoma patients from the University of Pennsylvania Health System
}

\begin{document}

% ----------------------------------------------------------------------------------------------------
% Title and abstract
% ----------------------------------------------------------------------------------------------------

%\title{\emph{GBMBench}: Towards Clincal Translation Of Glioblastoma Growth Models}

%\title{\emph{GLIO-MAP}: Glioblastoma Modeling for Adaptive Planning}

%\title{\emph{GLIMPACT}: Glioblastoma Model Performance and Clinical Translation}

%\title{\emph{GROWTHMAP}: Benchmarking and Translating Glioblastoma Growth Models for Personalized Therapy}

%\title{PREDICT-GBM: Platform for Robust Evaluation and Development of Individualized Computational Tumor Models in Glioblastoma}

\title{PREDICT-GBM: A multi-center platform to advance personalized glioblastoma radiotherapy planning}

\author[1]{Lucas Zimmer}
\author[1,2]{Jonas Weidner}
\author[3]{Michal Balcerak}
\author[3,4]{Florian Kofler}
\author[1]{Mara Krupa}
\author[5]{Ivan Ezhov}
\author[6,7]{Santiago Cepeda}
\author[8]{Ray Zirui Zhang}
\author[9,10]{John S. Lowengrub}
\author[3,11]{Bjoern Menze}
\author[1,2,11]{Benedikt Wiestler}

\affil[1]{AI for Image-Guided Diagnosis and Therapy, Technical University of Munich, Germany}
\affil[2]{Munich Center for Machine Learning (MCML)}
\affil[3]{Department of Quantitative Biomedicine, University of Zurich, Switzerland}
\affil[4]{Helmholtz AI, Helmholtz Zentrum München, Germany}
\affil[5]{AI in Healthcare and Medicine, Technical University of Munich, Munich, Germany}
\affil[6]{Department of Neurosurgery, Río Hortega University Hospital, Valladolid, Spain}
\affil[7]{Specialized Group in Biomedical Imaging and Computational Analysis (GEIBAC), Instituto de Investigación Biosanitaria de Valladolid (IBioVALL), Valladolid, Spain}
\affil[8]{Department of Mathematical Sciences, Worcester Polytechnic Institute, USA}
\affil[9]{Department of Mathematics, University of California, Irvine, USA}
\affil[10]{Department of Biomedical Engineering, University of California, Irvine, USA}
\affil[11]{BM and BW contributed equally as senior authors}

\correspondence{Correspondence: \texttt{zimmer.lucas.w@gmail.com}}

\abstracttext{Glioblastoma recurrence is largely driven by diffuse infiltration beyond radiologically visible tumor margins, yet standard radiotherapy, the mainstay of glioblastoma treatment, relies on uniform expansions that ignore patient-specific biological and anatomical factors. While computational models promise to map this invisible growth and guide personalized treatment planning, their clinical translation is hindered by the lack of standardized, large-scale benchmarking and reproducible validation workflows.
To bridge this gap, we present PREDICT-GBM, a comprehensive open-source platform that integrates a curated, longitudinal, multi-center dataset of 243 patients with a standardized evaluation pipeline, and fuels model development and validation. We demonstrate PREDICT-GBM's potential by training and benchmarking a novel U-Net–based recurrence prediction model against state-of-the-art biophysical and data-driven methods.
Our results show that both biophysical and deep-learning approaches significantly outperform standard-of-care protocols in predicting future recurrence sites while maintaining iso-volumetric treatment constraints. Notably, our U-Net model achieved a superior coverage of enhancing recurrence ($79.37 \pm 2.08\%$), markedly surpassing the standard-of-care (paired Wilcoxon signed-rank test, $p = 5.7 \times 10^{-6}$). Furthermore, the biophysical model GliODIL reached $78.91 \pm 2.08\%$ ($p = 4.5 \times 10^{-4}$), validating the platform’s ability to compare diverse modeling paradigms.
By providing the first rigorous, reproducible ecosystem for model training and validation, PREDICT-GBM eliminates a major bottleneck for personalized, computationally guided radiotherapy. This work establishes a new standard for developing computationally guided, personalized radiotherapy, with the platform, models, and data openly available at \url{github.com/BrainLesion/PredictGBM}.
}

\twocolumn[
  \maketitle
]

% ----------------------------------------------------------------------------------------------------
% Introduction
% ----------------------------------------------------------------------------------------------------
\section{Introduction}
%\vspace{-\baselineskip}
%\vspace{-0.08cm}
\label{sec:introduction}
\Ac{gbm} remains one of the greatest challenges in oncology, defying modern therapeutic advances, with median \ac{os} stagnating at 14-16 months since the introduction of adjuvant temozolomide to standard postoperative \ac{rt} in 2005~\cite{stupp2005radiotherapy, koshy2012improved, brown2022survival, mohammed2022survival, ostrom2018females, grochans2022epidemiology}. Glioblastoma is characterized by its highly invasive nature, with tumor cells diffusely infiltrating the surrounding brain parenchyma, leading to near-universal tumor recurrence despite aggressive, multimodal therapy~\cite{weller2013standards, Niyazi2023}. \Ac{soc} treatment consists of maximal safe resection followed by adjuvant \ac{rt} with concomitant and adjuvant chemotherapy~\cite{wen2020glioblastoma}. MRI is central to both treatment planning and longitudinal disease monitoring. Because MRI only delineates regions of high tumor cell density, and even advanced MRI struggles to delineate the diffuse infiltration of tumor cells into the surrounding parenchyma~\cite{lasocki2019non}, microscopic tumor infiltration is addressed in clinical practice by applying a uniform margin to the visible tumor to define the \ac{ctv} for irradiation, in accordance with current \ac{rt} guidelines~\cite {Niyazi2023}. However, this margin-based approach does not capture patient-specific tumor biology, invasion patterns, or anatomy. Consequently, this "one-size-fits-all" approach risks simultaneously undertreating the tumor and overtreating the surrounding brain.

\begin{table*}[ht]
\centering
\small
\begin{tabular}{@{}lllll@{}}
\toprule
\textbf{Patient cohort}                                    & \textbf{All patients} & \textbf{RHUH}      & \textbf{LUMIERE}  & \textbf{TUM-GBM}  \\ \midrule
\textbf{N}                                                 & 243                   & 40                 & 61                & 142               \\
\textbf{Tumor volume ($\boldsymbol{\mathrm{cm}^3}$)}       &                       &                    &                   &                   \\
\quad Average/stdev                                        & $32.3 \pm 26.2$       & $35.9 \pm 27.1$    & $33.8 \pm 25.0$   & $30.7 \pm 26.4$   \\
\quad Median                                               & 27.5                  & 31.4               & 27.5              & 25.8              \\
\quad IQR                                                  & {[}12.2 - 47.3{]}     & {[}12.3 - 52.9{]}  & {[}13.9 - 49.8{]} & {[}9.2 - 45.2{]}  \\
\textbf{Recurrence volume ($\boldsymbol{\mathrm{cm}^3}$)}  &                       &                    &                   &                   \\
\quad Average/stdev                                        & $13.7 \pm 19.7$       & $22.9 \pm 24.0$    & $20.3 \pm 26.7$   & $8.6 \pm 11.9$    \\
\quad Median                                               & 5.6                   & 15.0               & 8.5               & 4.1               \\
\quad IQR                                                  & {[}1.7-17.7{]}        & {[}5.6 - 31.9{]}   & {[}2.0 - 26.7{]}  & {[}1.3 - 10.2{]}   \\
\textbf{Center of Mass Distance (cm)}                      &                       &                    &                   &                   \\
\quad Average/stdev                                        & $2.2 \pm 1.9$         & $1.7 \pm 1.5$      & $3.0 \pm 2.8$     & $1.9 \pm 1.4$     \\
\quad Median                                               & 1.6                   & 1.4                & 2.1               & 1.6               \\
\quad IQR                                                  & {[}0.9 - 2.6{]}       & {[}0.8 - 2.1{]}    & {[}1.2 - 3.4{]}   & {[}0.9 - 2.5{]}   \\
\textbf{Age (years)}                                       &                       &                    &                   &                   \\
\quad Average/stdev                                        & $62.4 \pm 10.2$       & $63.0 \pm 9.2$     & $62.1 \pm 9.6$    & $62.4 \pm 10.8$   \\
\textbf{OS (months)}                                       &                       &                    &                   &                   \\
\quad Average/stdev                                        & $16.6 \pm 9.3$        & $14.4 \pm 8.9$     & $20.9 \pm 10.1$   & $15.4 \pm 8.6$    \\
\textbf{Diagnosis}                                         &                       &                    &                   &                   \\
\quad WHO grade 4 glioma                                   & $100$~\%              & $100$~\%           & $100$~\%          & $100$~\%          \\
\quad IDH Wildtype                                         & $98.4$~\%             & $90$~\%            & $100$~\%          & $100$~\%          \\
\quad IDH Mutant                                           & $1.6$~\%              & $10.0$~\%          & $0.0$~\%          & $0.0$~\%          \\
\bottomrule
\end{tabular}
\caption{\textbf{Patient characteristics}}
\label{tab:patient_characteristics}
\end{table*}

To enable the true personalization of \ac{rt} volumes, computational models of glioblastoma have been proposed. These models simulate tumor growth and thereby generate three-dimensional tumor cell density distributions beyond the margins visible in standard MRI, with the potential to support biologically informed, patient-specific \ac{rt} target definition. Early approaches predominantly relied on reaction–diffusion models of tumor cell proliferation and migration, solved using PDE-constrained optimization frameworks incorporating Bayesian inference~\cite{menze2011image, lipkova2019personalized, konukoglu2009image, frieboes2007computer}, numerical solvers~\cite{mang2012biophysical, scheufele2020image, scheufele2020fully, subramanian2020multiatlas, subramanian2022ensemble}, or analytical approaches~\cite{unkelbach2014radiotherapy}. High computational cost and limited availability of suitable longitudinal imaging data have historically restricted these studies to small patient cohorts. Subsequent work leveraged the advances of deep learning for data-driven approaches either by learning the forward problem to optimize a surrogate model~\cite{ezhov2021geometry, viguerie2022data, martens2022deep} or by learning the inverse problem, mapping from images to growth parameters in the PDE~\cite{ezhov2019neural, pati2021estimating, ezhov2023learn}. Recent work further combines model-driven and data-driven principles by learning a prior for the growth parameters before utilizing evolutionary sampling with a numerical solver~\cite{weidner2024learnable}, introducing a neural PDE solver by performing gradient-based optimization with respect to the growth parameters during inference~\cite{weidner2024rapid}, or introducing a physics term in the loss function~\cite{zhang2025personalized, balcerak2025individualizing, balcerak2024physics, weidner2024spatial, metz2023towards}. Another recent line of work frames the problem as direct recurrence prediction, bypassing explicit estimation of a tumor cell density map and instead relying on purely data-driven learning approaches~\cite{cepeda2023predicting, cepeda2025radiomics, tran2025novel}.

Despite these promising methodological advances, the clinical translation of glioblastoma growth models remains limited. Successful translation requires comparative validation across sufficiently large and heterogeneous patient cohorts; however, prior research has predominantly relied on small, institution-specific datasets without standardized evaluation protocols. Only a limited number of studies report evaluations on cohorts exceeding 100 patients~\cite{subramanian2022ensemble, weidner2024spatial, balcerak2025individualizing, scheufele2020fully}. This fragmentation stems largely from practical barriers to data access. Robust assessment of growth models necessitates longitudinal imaging, specifically, pre-treatment MRI and follow-up scans capturing the first tumor recurrence, which poses significant curation challenges. Furthermore, sharing such high-resolution cranial MRI data is often restricted by privacy and anonymization concerns. While platforms like \ac{tcia}~\cite{Clark2023TCIA} provide valuable access to de-identified glioblastoma datasets, few publicly available collections meet the specific longitudinal criteria required for growth modeling. Consequently, the field lacks a standardized benchmark, rendering meaningful comparison between distinct modeling approaches impossible.

In addition to data scarcity, the lack of a standardized processing pipeline introduces significant variability. Most models require co-registered, multi-sequence MRI inputs, along with segmentations of both tumor and healthy brain tissue. While the multimodal \ac{brats} benchmark has established standards for tumor segmentation~\cite{menze2014multimodal, kofler2020brats, kofler2025brats, de20242024, kofler2025brainlesion}, extracting the healthy tissue geometry required to model cell migration remains an open challenge. General-purpose neuroimaging tools, such as FreeSurfer and SynthSeg~\cite{fischl2012freesurfer, billot2023synthseg}, FastSurfer~\cite{henschel2020fastsurfer}, FSL FAST~\cite{zhang2001segmentation}, or ANTs~\cite{tustison_antsx_2021}, offer solutions for healthy brains but often struggle in the presence of significant pathology. Specifically, accurate tissue classification in the peritumoral region is confounded by mass effect, edema, and postoperative changes. As a result, researchers often rely on custom, unstandardized preprocessing workflows.
Lastly, there is no consensus on the evaluation metrics. This leads to slightly different definitions of, e.g., "recurrence coverage" and related concepts across studies, effectively precluding the community from objectively comparing different approaches to modeling and predicting tumor growth for personalized therapy. These disparities in data curation, image processing, and evaluation metrics highlight the critical need for an open, extensible platform to unify and accelerate model development.

To address these challenges, we introduce PREDICT-GBM: the first open, extensible platform for glioblastoma growth models and recurrence prediction models, designed to enable standardized evaluation and facilitate their translation to clinical application. PREDICT-GBM provides a large, curated, and fully processed longitudinal glioblastoma imaging dataset, together with a unified evaluation pipeline that enables head-to-head comparison of state-of-the-art computational models against each other and against the current clinical \ac{soc}, and provides a critical resource for the development and evaluation of new approaches to personalized glioblastoma treatment. By effectively comparing diverse, modeling- and learning-based approaches, PREDICT-GBM provides critical insights into the state of the field and future research directions. The entire framework is fully open source, allowing researchers to process additional datasets, develop and benchmark new growth models, and perform reproducible evaluations in a simple, standardized manner.

% ----------------------------------------------------------------------------------------------------
% Results
% ----------------------------------------------------------------------------------------------------
\section{Results}
\label{sec:results}

\begin{table*}[t]
\centering
\begin{subtable}[t]{\textwidth}
\centering
\begingroup
\scriptsize
\setlength{\tabcolsep}{3pt} 
\begin{tabular}{@{\hspace{.1cm}}l@{\hspace{.25cm}}c@{\hspace{.25cm}}c@{\hspace{.25cm}}c@{\hspace{.25cm}}c@{\hspace{.25cm}}c@{\hspace{.25cm}}c@{\hspace{.25cm}}c@{\hspace{.25cm}}c@{\hspace{.1cm}}}
\toprule
\textbf{Dataset}         & \textbf{SOC}          & \textbf{LOTI}                 & \textbf{GliODIL}             & \textbf{U-Net}              & \textbf{PINN-GBM}     & \textbf{LMI}          & \textbf{nnU-Net}      & \textbf{GlioMap}    \\ \midrule
RHUH                     & $82.64 \pm 4.26$      & $82.38 \pm 3.97$              & $83.64 \pm 4.16$             & $\mathbf{85.26 \pm 3.97}$   & $83.17 \pm 4.17$      & $70.31 \pm 4.00$      & $53.69 \pm 5.12$      & $-$                 \\
TUM-GBM                  & $79.89 \pm 2.81$      & $81.88 \pm 2.58$              & $\mathbf{82.13 \pm 2.61}$*   & $81.80 \pm 2.66$*           & $80.84 \pm 2.78$      & $68.48 \pm 2.92$      & $60.57 \pm 2.71$      & $-$                 \\
LUMIERE                  & $67.91 \pm 4.76$      & $68.26 \pm 4.56$              & $68.29 \pm 4.70$             & $\mathbf{69.87 \pm 4.72}$*  & $67.98 \pm 4.78$      & $62.31 \pm 4.62$      & $59.29 \pm 4.89$      & $-$                 \\
DIFFUSION                & $82.73 \pm 2.84$      & $\mathbf{86.12 \pm 2.40}$*    & $85.38 \pm 2.63$*            & $84.63 \pm 2.70$*           & $83.41 \pm 2.86$      & $71.37 \pm 3.04$      & $62.51 \pm 3.15$      & $83.73 \pm 2.53$    \\ 
PREDICT-GBM              & $77.34 \pm 2.17$      & $78.54 \pm 2.03$              & $78.91 \pm 2.08$*            & $\mathbf{79.37 \pm 2.08}$*  & $77.99 \pm 2.16$      & $67.23 \pm 2.10$      & $59.12 \pm 2.45$      & $-$                 \\ \midrule 
$p_{\mathrm{PREDICT}}$   & $-$                   & $0.37$                        & $4.5 \times 10^{-4}$         & $5.7 \times 10^{-6}$        & $0.057$               & $1.00$                & $1.00$                & $-$
\end{tabular}
\caption{Enhancing recurrence}
\label{tab:rec_coverage_enhancing}
\endgroup
\end{subtable}

\vspace{2em}

\begin{subtable}[t]{\textwidth}
\centering
\begingroup
\scriptsize
\setlength{\tabcolsep}{3pt} 
\begin{tabular}{@{\hspace{.1cm}}l@{\hspace{.25cm}}c@{\hspace{.25cm}}c@{\hspace{.25cm}}c@{\hspace{.25cm}}c@{\hspace{.25cm}}c@{\hspace{.25cm}}c@{\hspace{.25cm}}c@{\hspace{.25cm}}c@{\hspace{.1cm}}}
\toprule
\textbf{Dataset}         & \textbf{SOC}          & \textbf{LOTI}                  & \textbf{GliODIL}            & \textbf{U-Net}               & \textbf{PINN-GBM}     & \textbf{LMI}          & \textbf{nnU-Net}      & \textbf{GlioMap}      \\ \midrule
RHUH                     & $82.93 \pm 4.25$      & $82.76 \pm 3.97$               & $83.99 \pm 4.15$            & $\mathbf{85.54 \pm 3.97}$    & $83.44 \pm 4.17$      & $70.60 \pm 4.02$      & $53.91 \pm 5.13$      & $-$                   \\
TUM-GBM                  & $79.93 \pm 2.80$      & $82.04 \pm 2.57$               & $\mathbf{82.27 \pm 2.61}$*  & $81.90 \pm 2.65$*            & $81.01 \pm 2.78$      & $68.35 \pm 2.91$      & $60.69 \pm 2.67$      & $-$                   \\
LUMIERE                  & $67.78 \pm 4.79$      & $68.17 \pm 4.59$               & $68.15 \pm 4.74$            & $\mathbf{69.74 \pm 4.76}$*   & $67.81 \pm 4.81$      & $62.23 \pm 4.65$      & $59.01 \pm 4.91$      & $-$                   \\
DIFFUSION                & $82.75 \pm 2.83$      & $\mathbf{86.26 \pm 2.39}$*     & $85.52 \pm 2.62$*           & $84.73 \pm 2.69$*            & $83.59 \pm 2.85$      & $71.38 \pm 3.03$      & $62.53 \pm 3.14$      & $83.76 \pm 2.52$      \\
PREDICT-GBM              & $77.37 \pm 2.17$      & $78.68 \pm 2.03$               & $79.01 \pm 2.08$*           & $\mathbf{79.44 \pm 2.09}$*   & $78.09 \pm 2.16$*     & $67.18 \pm 2.17$      & $59.15 \pm 2.45$      & $-$                   \\ \midrule 
$p_{\mathrm{PREDICT}}$   & $-$                   & $0.33$                         & $3.3 \times 10^{-4}$        & $5.4 \times 10^{-6}$         & $0.049$               & $1.00$                & $1.00$
\end{tabular}
\caption{Recurrence core (enhancing, necrotic)}
\label{tab:rec_coverage_enhancing_necrosis}
\endgroup
\end{subtable}

\caption{\textbf{Coverage and standard error for (a) enhancing recurrence, and (b) recurrence core (enhancing + necrotic).} * indicates $p<0.05$ (paired Wilcoxon signed-rank test) comparing model plans to standard plans (SOC~\cite{Niyazi2023}); \textbf{bold} indicates best value per dataset. $p_{\mathrm{PREDICT}}$: comparison against \ac{soc} in the full PREDICT-GBM dataset ($N=243$). DIFFUSION: subset of PREDICT-GBM with ADC/DTI images ($N=112$).}
\label{tab:combined}
\end{table*}

\paragraph{Patient characteristics}
The combined PREDICT-GBM cohort included 243 subjects, curated from three datasets (LUMIERE~\cite{suter2022lumiere}, RHUH~\cite{cepeda2023rio}, and TUM-GBM) that met our data inclusion criteria. Patient characteristics are summarized in Table~\ref{tab:patient_characteristics}. The median baseline tumor volume was $27.5\,\mathrm{cm}^{3}$, while the median recurrence volume was $5.6\,\mathrm{cm}^{3}$, with cases spanning a broad range of lesion sizes across the cohort, well in line with the broad heterogeneity seen in clinical routine. To contextualize spatial patterns, Table~\ref{tab:patient_characteristics} also reports the distance between the primary tumor and the recurrence, measured between their centers of mass (CoM). Recurrences were predominantly local, a well-known finding in glioblastoma~\cite{chamberlain2011radiographic}, with a median \ac{com} distance of $1.6\,\mathrm{cm}$, while the cohort also included more distant recurrences with \ac{com} displacements of up to $9.1\,\mathrm{cm}$.

\begin{figure}[t]
\centering
\includegraphics[width=\linewidth]{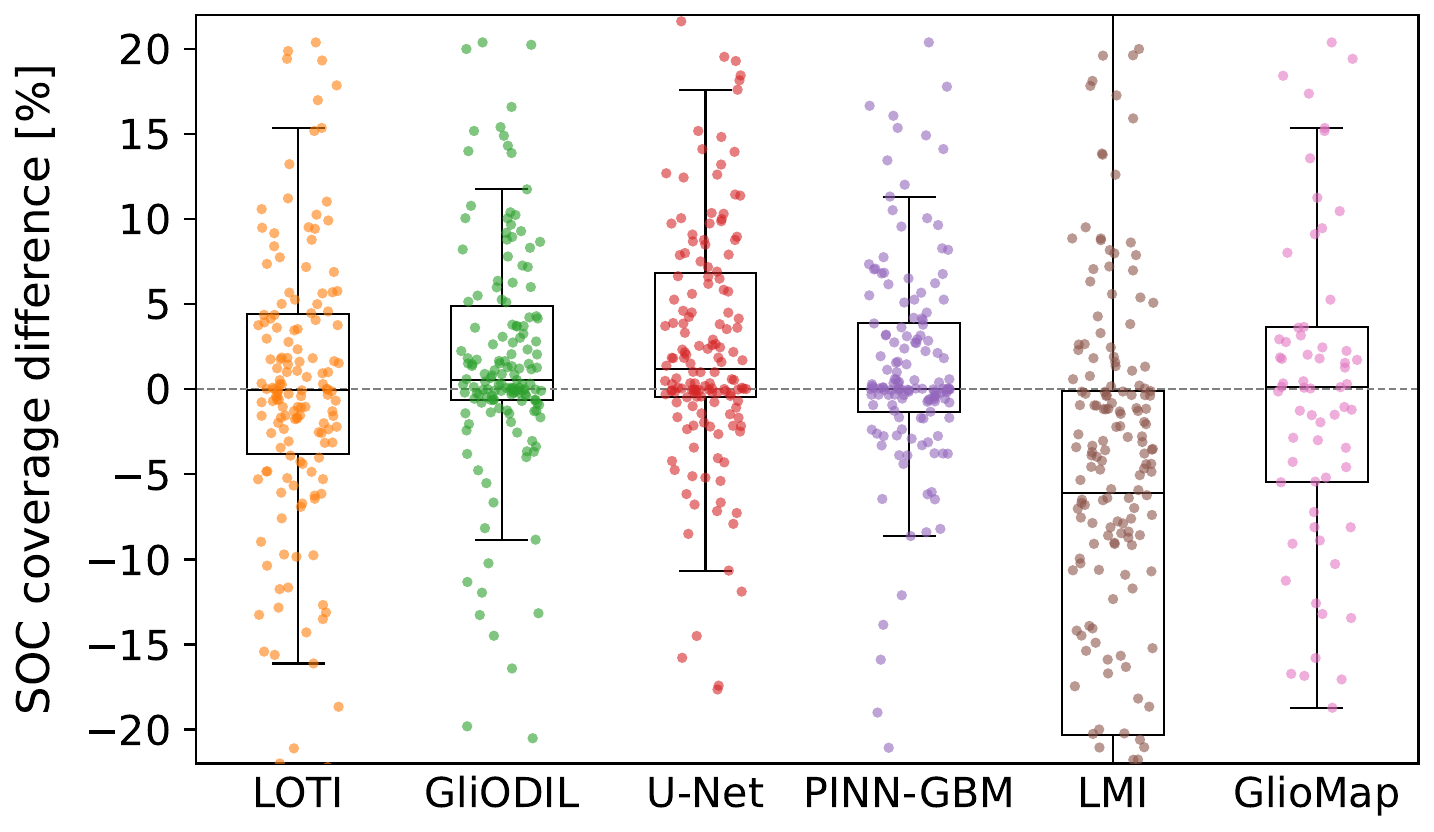}
\caption{Difference between model coverage and \ac{soc} coverage, excluding cases in which model and \ac{soc} achieved the exact same coverage. Positive values indicate a higher model coverage compared to \ac{soc}.}
\label{fig:coverage_diff}
\end{figure}

\begin{figure*}[t]
\centering
\includegraphics[width=0.85\linewidth]{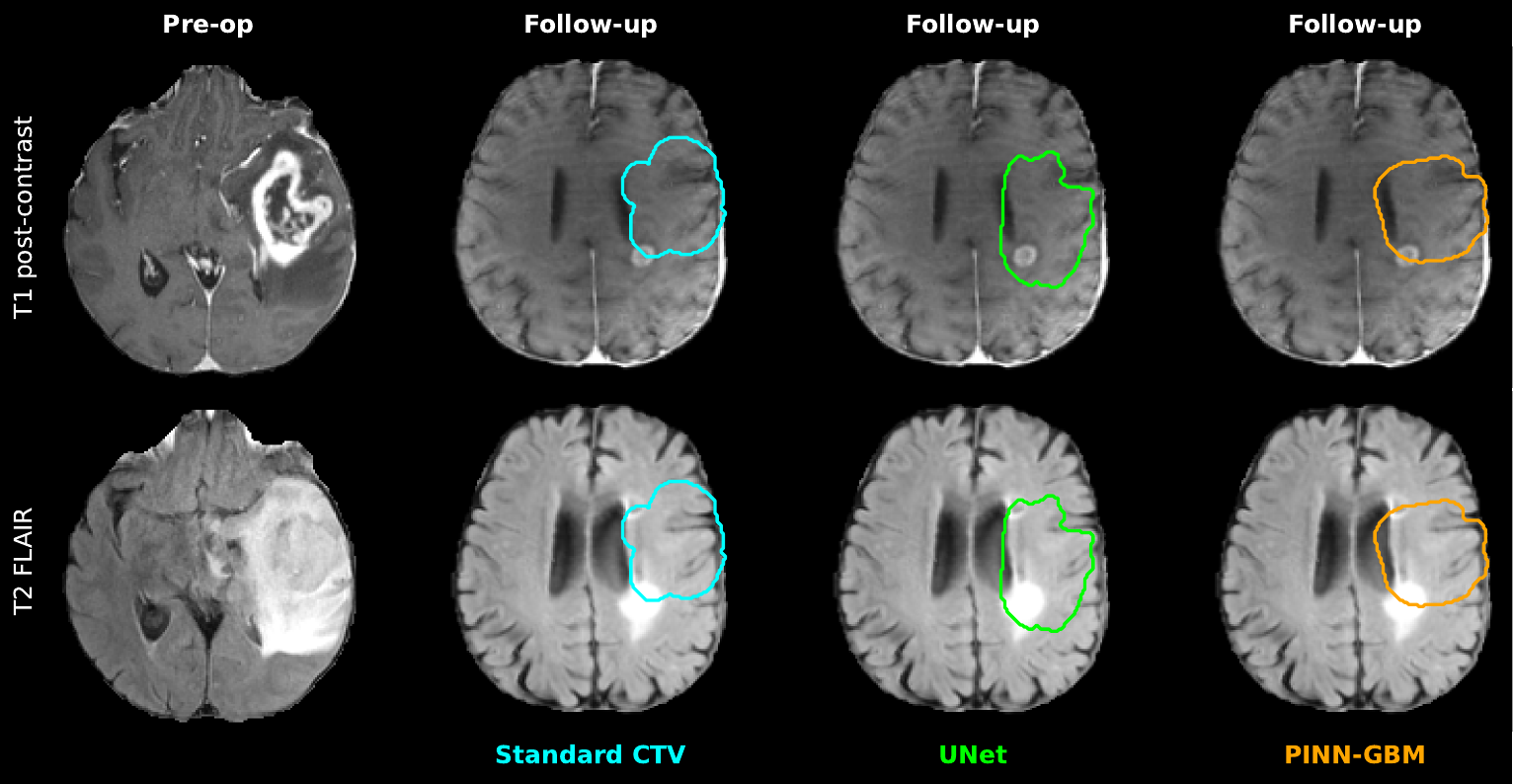}

\vspace{2mm}

\includegraphics[width=0.85\linewidth]{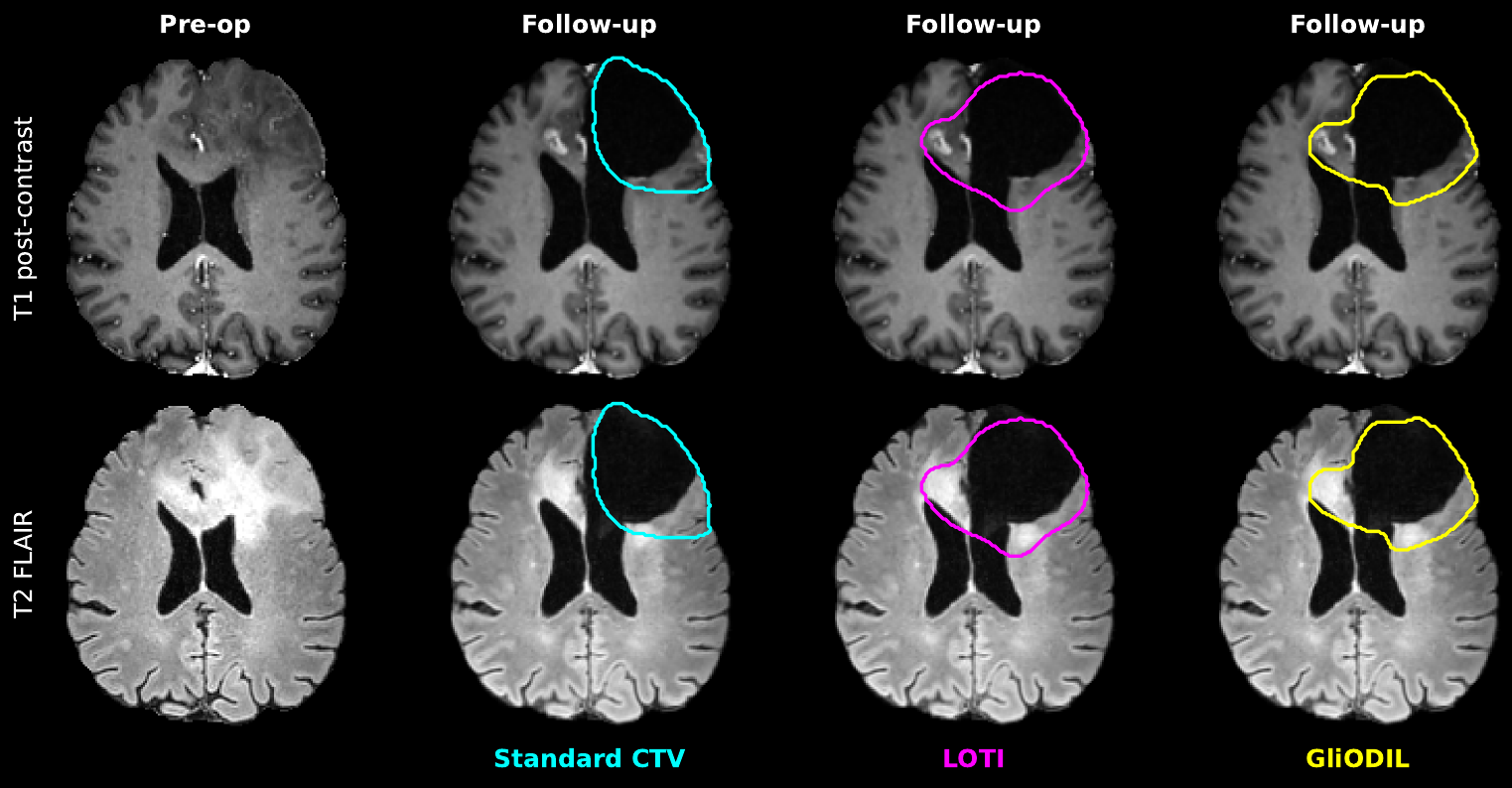}
\caption{\textbf{Comparison of CTVs for 2 example patients.} Axial slices are centered on the pre-operative tumor center of mass for pre-op and on the recurrence center of mass for follow-up exams. Follow-up images are deformably registered to the pre-operative space. \ac{ctv}s, shown as contours, are iso-volumetric: \ac{soc} \ac{ctv} (cyan), U-Net \ac{ctv} (lime), PINN-GBM \ac{ctv} (orange), \ac{loti} \ac{ctv} (magenta), and GliODIL \ac{ctv} (yellow). While the \ac{soc} \ac{ctv} misses significant portions of the recurrent tumor, the model-based \ac{ctv}s redistribute volume to cover the recurrence, while simultaneously sparing normal brain tissue.}
\label{fig:qualitative}
\end{figure*}

\paragraph{Recurrence coverage}
Model performance was evaluated for a wide variety of models, spanning both learning- and modeling-based approaches as well as their combination, using two clinically relevant recurrence definitions: (i) enhancing recurrence (Table~\ref{tab:rec_coverage_enhancing}), and (ii) enhancing plus necrotic recurrence (Table~\ref{tab:rec_coverage_enhancing_necrosis}). Results are reported for each individual contributing dataset and for the aggregated PREDICT-GBM cohort, with representative qualitative examples shown in Fig.~\ref{fig:qualitative}. We also report results for the subset of patients with available diffusion images, as this modality is required for GlioMap (excluding RHUH, which was included in the GlioMap training). Please note that all plans are \textit{iso-volumetric} to the standard plan to allow for an unbiased, fair comparison. Across the combined cohort, our newly developed U-Net achieved the highest mean coverage for both recurrence definitions, closely followed by GliODIL. The U-Net, GliODIL, \ac{loti}, PINN-GBM, and GlioMap all performed at least as well as the standard plan. In contrast, both \ac{lmi} and the nnU-Net baseline underperformed by approximately 10\% and 20\%, respectively, across the aggregated cohort. Including necrosis in the recurrence definition produced only modest changes in mean coverage compared with an enhancing-only evaluation.

\begin{figure}[t]
\centering
\includegraphics[width=\linewidth]{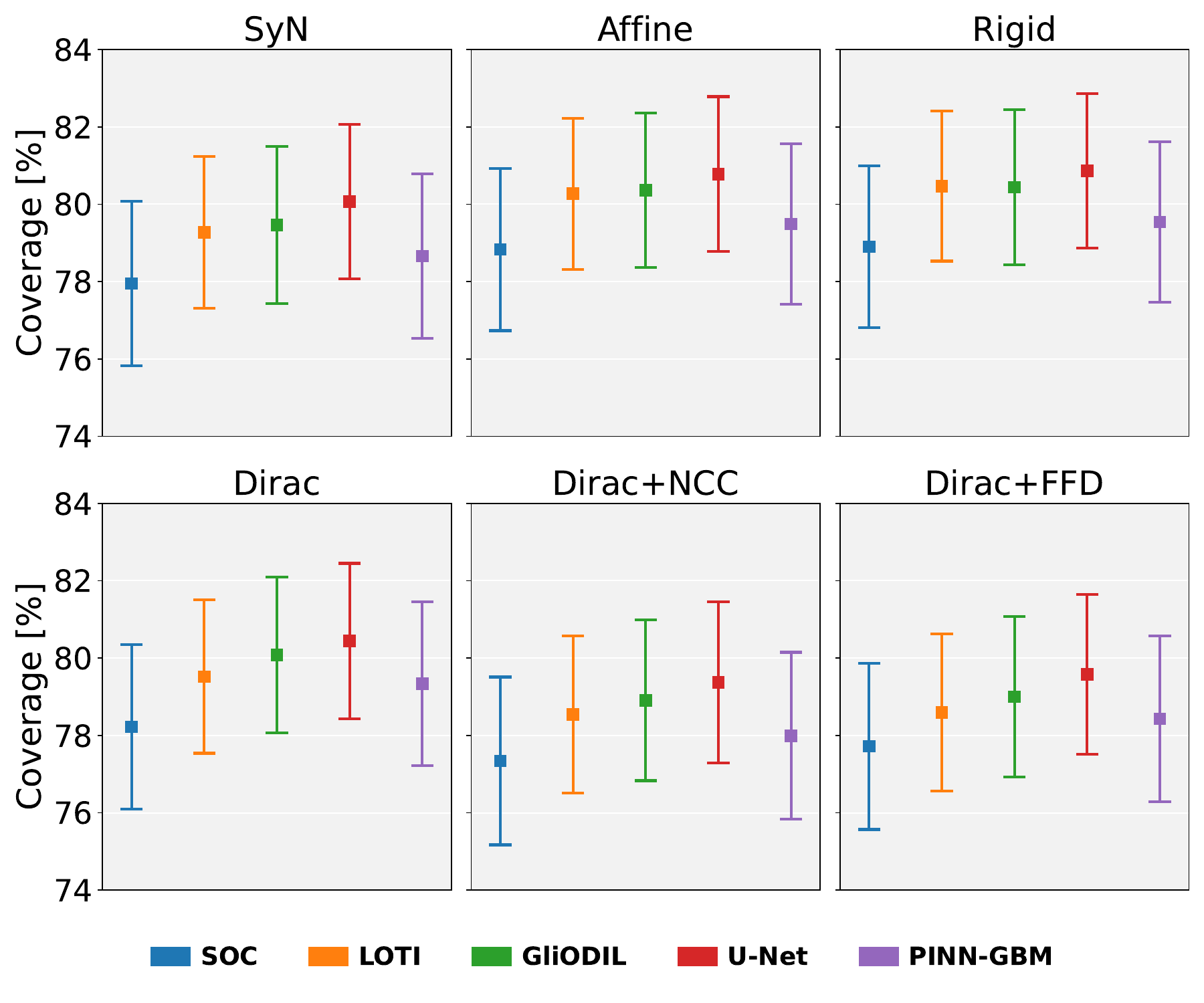}
\caption{Coverage of enhancing recurrence for the combined cohort using different longitudinal registration algorithms to map the follow-up exam from atlas space to the pre-treatment exam. Error bars denote the standard error of the mean.}
\label{fig:longitudinal_registration}
\end{figure}

\paragraph{Statistical analysis}
Both our U-Net and GliODIL showed significant improvements in enhancing-only and enhancing + necrosis recurrence ($p<10^{-5}$ and $p<10^{-3}$, respectively). PINN-GBM achieved $p$-values close to 0.05, with $p<0.05$ for the enhancing+necrotic recurrence. At the dataset level, our U-Net achieved significant ($p<0.05$) improvement over the standard plan on all datasets except RHUH. GliODIL achieved significant improvement over SOC on TUM-GBM and, with \ac{loti}, on the diffusion subset. To further investigate spatial factors associated with model performance, we analyzed the relationship between recurrence location and the improvement in coverage over the standard \ac{ctv}. Using Spearman rank correlation $\rho$, we observed a weak positive correlation ($\rho$ between 0.2 and 0.3) between \ac{com} distance and improvement over the standard \ac{soc} within the clinically relevant \ac{com} distance range of $15-30\,\mathrm{mm}$ for our U-Net, GliODIL and \ac{loti}. This suggests that model-based \ac{soc}s tend to provide greater benefit for recurrences occurring at intermediate distances from the primary tumor, a spatial range that may be particularly challenging for uniform margin-based expansions. Figure~\ref{fig:coverage_diff} illustrates the per-model improvement in recurrence coverage relative to \ac{soc}. The U-Net demonstrated the highest median improvement, followed by GliODIL. GliODIL and PINN-GBM demonstrated lower IQR, indicating greater robustness and more consistent patient-level performance.

\begin{figure}[t]
\includegraphics[width=\linewidth]{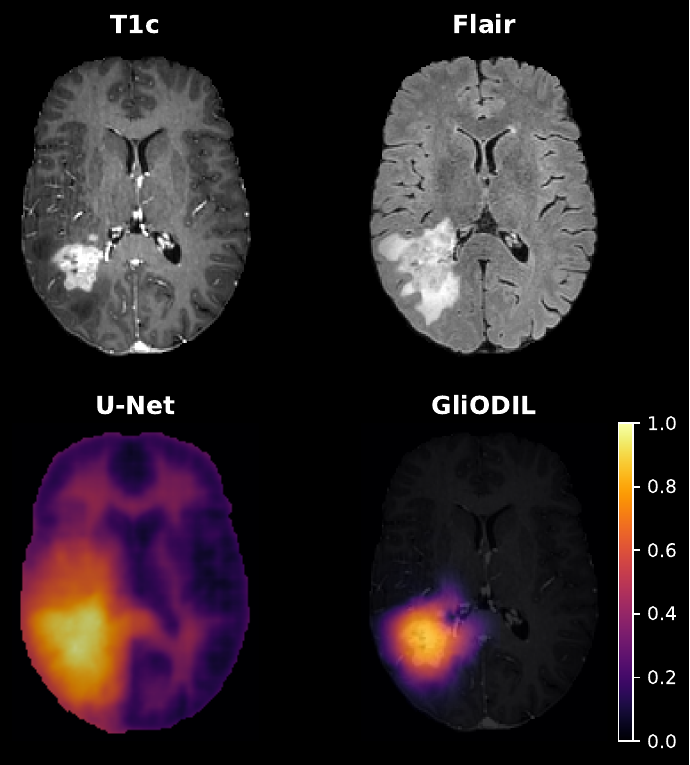}
\caption{Representative outputs of our U-Net and GliODIL, showing predicted probability of future recurrence and estimated tumor cell concentration, respectively.}
\label{fig:supp:unet:pred}
\end{figure}

\paragraph{Longitudinal registration}
Accurate evaluation of recurrence coverage requires voxel-wise correspondence between the pre-treatment and follow-up exams. Longitudinal alignment in glioblastoma is challenging due to mass effect, postoperative changes, and therapy-related anatomical deformation. To assess the sensitivity of our results to the choice of longitudinal registration, we compared rigid and affine registration with deformable ANTs \ac{syn}~\cite{avants2008symmetric} as well as the 2022 MICCAI BraTSReg challenge winning algorithm (Dirac+NCC)~\cite{baheti2021brain,mok2022unsupervised,mok2022robust} and an alternative instance optimization (Dirac+FFD) for mapping the follow-up exam into the pre-operative reference space. Figure~\ref{fig:longitudinal_registration} reports enhancing-recurrence coverage for each registration method. While absolute coverage values varied slightly across algorithms, the relative ranking of planning approaches remained unchanged. Consistently, conclusions from paired Wilcoxon signed-rank tests were stable across registration choices, indicating that the reported statistical findings are not driven by the specific longitudinal registration method, but are robust.

\paragraph{U-Net predictions}
An example prediction is illustrated in Figure~\ref{fig:supp:unet:pred}. Qualitative analysis indicates that white matter regions are assigned a higher probability of recurrence, while gray matter and CSF are assigned a lower probability. This is consistent with the higher diffusivity in white matter and with biological and clinical studies on glioblastoma recurrence~\cite{brooks2021white, sansone2023patterns}. Additionally, cell migration from one hemisphere to the other is predicted mostly through the corpus callosum, indicating that the model does not only predict via proximity but indeed learns meaningful, biologically plausible spatial patterns.

\begin{figure*}[t]
\centering
\includegraphics[width=0.8\linewidth]{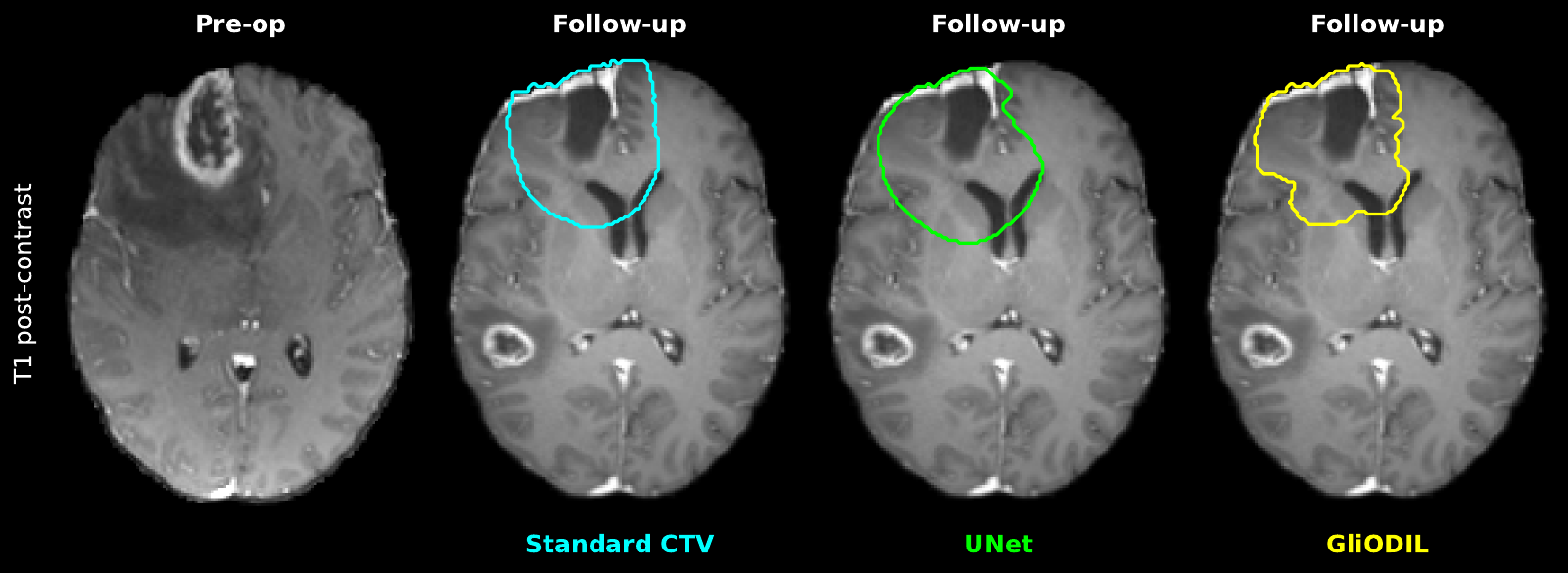}

\vspace{2mm}

\includegraphics[width=0.8\linewidth]{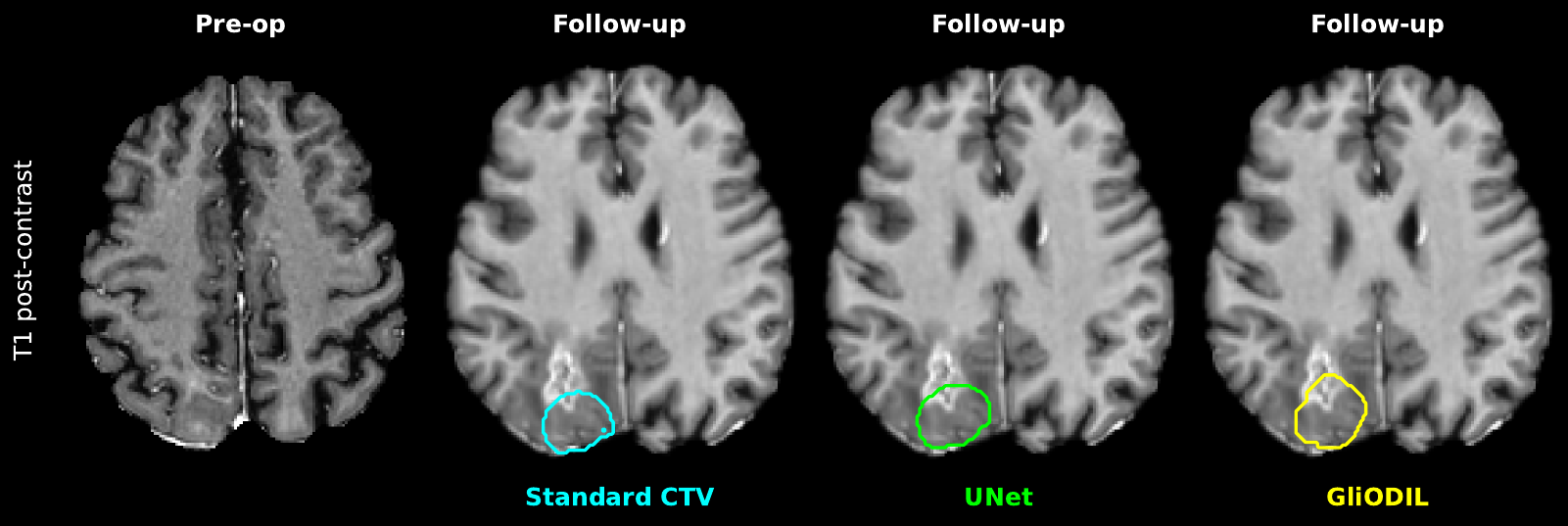}

\vspace{2mm}

\includegraphics[width=0.8\linewidth]{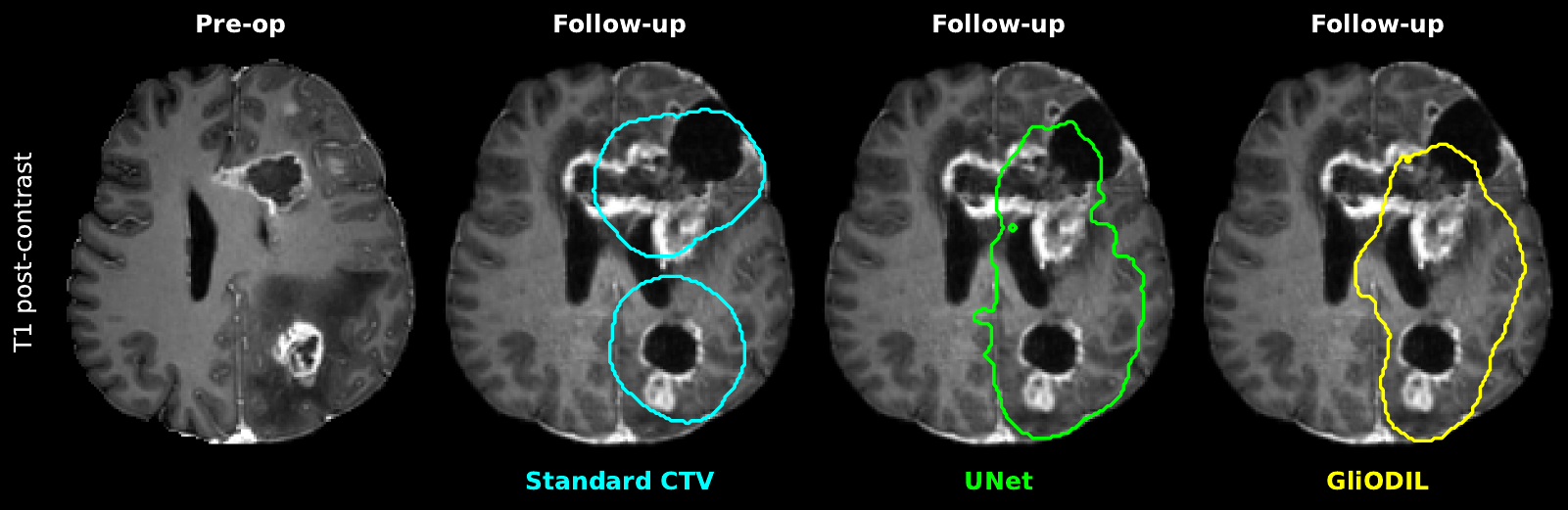}

\caption{\textbf{Comparison of CTVs for failure cases.} Axial slices are centered on the pre-operative tumor center of mass for pre-op and on the recurrence center of mass for follow-up exams. Follow-up images are deformably registered to the pre-operative space. \ac{ctv}s, shown as contours, are iso-volumetric: \ac{soc} \ac{ctv} (cyan), U-Net \ac{ctv} (lime), and GliODIL \ac{ctv} (yellow). The top case illustrates a distant recurrence, the middle case a small primary tumor with a comparatively large recurrence, and the bottom case a multifocal tumor.}
\label{fig:failure_cases}
\end{figure*}

\paragraph{Failure cases}
Representative cases with particularly low recurrence coverage are shown in Figure~\ref{fig:failure_cases}. Distant recurrences were not adequately captured by either the \ac{ctv} or the model-based \ac{ctv}s. In cases with very small initial tumors, the limited baseline volume constrains iso-volumetric redistribution, resulting in near-spherical \ac{ctv}s that fail to extend toward the eventual recurrence site. For multifocal tumors, the \ac{soc} outperformed model-based approaches, as physics-informed models typically assume a single initial tumor position and therefore do not explicitly account for multifocal disease, whereas data-driven models suffer from underrepresentation of multifocal tumors in the training data.

% ----------------------------------------------------------------------------------------------------
% Discussion
% ----------------------------------------------------------------------------------------------------
\section{Discussion}
\label{sec:discussion}
% Paragraph 1 (Motivation, Contribution)
Although advances in radiotherapy delivery have enabled highly conformal dose distributions, current glioblastoma target definition still largely relies on isotropic margin expansions around MRI-visible disease, without accounting for spatially heterogeneous invasion patterns, patient-specific tumor biology, or anatomical constraints. This paradigm risks undertreating infiltrative tumor that extends beyond the chosen margin, as shown by previous research~\cite{niyazi2023estro, unkelbach2014radiotherapy, menze2011image, konukoglu2009image, frieboes2007computer}, while simultaneously irradiating substantial volumes of normal brain tissue, with negative implications for cognitive function, quality of life, and even survival~\cite{soffietti2023neurotoxicity}. Computational growth models offer an appealing alternative: by estimating tumor cell density beyond the enhancing lesion and edema, they could support biologically informed and individualized target volumes. Furthermore, purely data-driven models could leverage advancements in machine learning for recurrence prediction, a task conceptually distinct from glioblastoma growth modeling and that does not explicitly estimate underlying tumor cell dynamics. However, progress toward clinical translation has been constrained by limited access to suitable longitudinal imaging, heterogeneous preprocessing choices, and inconsistent evaluation protocols, making it difficult to determine when model-based plans truly improve upon standard practice. This study addresses these barriers by introducing PREDICT-GBM as a standardized platform for glioblastoma growth modeling in the context of radiotherapy planning. PREDICT-GBM provides (i) a large, expert-curated, longitudinal multi-center dataset, (ii) an end-to-end open-source pipeline spanning preprocessing, segmentation, registration, model inference, and plan evaluation, and (iii) a containerized collection of growth and data-driven models that can be readily executed within the pipeline and easily extended to incorporate new modeling approaches. By providing both processed inputs and an extensible framework, PREDICT-GBM is intended to reduce methodological variability and shift the emphasis toward clinically relevant comparisons across heterogeneous cohorts, which are critical for future clinical translation. Importantly, our results demonstrate that both data-driven recurrence prediction models and biophysical growth models can achieve higher coverage of future recurrence sites for glioblastoma radiotherapy planning. This stringent evaluation of parallel approaches allows us to draw meaningful conclusions and outline promising future research directions: (i) expanding and enriching clinical datasets through multi-center data sharing and the inclusion of spatially resolved biopsy data to improve biological calibration and validation and to improve performance of data-driven models (ii) advancing physics-driven modeling beyond classical diffusion–reaction formulations by incorporating mechanisms such as mass effect, multi-compartment tumor representations, and explicit handling of multifocal disease, as well as integrating additional modalities including diffusion, perfusion, or PET imaging, and (iii) moving beyond purely geometric target comparisons toward dose-aware evaluation frameworks, where model-derived risk maps can guide dose painting or boosting while respecting organ-at-risk constraints, ultimately paving the way to clinical adoption of computational models for individualizing radiotherapy planning.

% Paragraph 2 (Results)
Across the aggregated cohort, for both the enhancing and enhancing-plus-necrotic recurrence definitions, our U-Net achieved the highest mean recurrence coverage, followed by GliODIL, \ac{loti} and PINN-GBM, each of which outperformed the \ac{soc} target definition. GlioMap likewise exceeded \ac{soc} on the subset of patients with available diffusion imaging. \ac{lmi} (trained on synthetic data) and the nnU-Net baseline produced lower mean coverage than \ac{soc} planning. Importantly, the U-Net and GliODIL demonstrated a statistically significant improvement over the standard plan for both enhancing-only recurrence and enhancing-plus-necrotic recurrence in the pooled cohort (paired Wilcoxon signed-rank tests). Our U-Net exhibited the largest median improvement while GliODIL showed a narrower interquartile range, indicating more consistent patient-level benefit with fewer outliers. These results suggest that growth models incorporating biophysical structure may confer greater robustness than purely data-driven recurrence prediction under the dataset sizes and domain shifts that currently characterize publicly accessible longitudinal \ac{gbm} imaging. At the same time, the absence of a consistent pooled-cohort significance advantage for other models over standard planning for enhancing-only definitions underscores that mean coverage improvements alone do not guarantee consistent patient-level benefit. Adding necrosis to the enhancing recurrence definition produced only modest changes in mean coverage, consistent with necrosis often co-localizing with enhancing progression and therefore not substantially altering the spatial evaluation target. Because model-based targets were constrained to be iso-volumetric with the \ac{soc} plan, any increase in recurrence coverage necessarily corresponds to an increase in specificity. Therefore, our results demonstrate that, in our setting, model-based radiation plans can increase coverage of regions of later recurrence while simultaneously reducing irradiation of healthy brain tissue, potentially leading to increased survival and lower treatment-related toxicity if adopted in clinical practice.

Further analysis of spatial patterns revealed a weak positive correlation between \ac{com} distance and coverage improvement over the standard \ac{ctv} within the clinically relevant $15-30\,\mathrm{mm}$ range. This finding suggests that model-based approaches may be particularly beneficial for intermediate distances, beyond the typical high-density tumor core but within a biologically plausible infiltration range, where uniform margin expansions may be suboptimal. Finally, sensitivity analyses of the longitudinal registration step demonstrated that, while absolute coverage values varied slightly across registration algorithms, the relative ranking of methods remained unchanged, and statistical significance conclusions were robust across registration strategies.

% Paragraph 3 (Key challenge - preprocessing)
A key challenge in comparative evaluation for recurrence prediction is the requirement for high-quality, disease-specific preprocessing. Model performance depends on upstream preprocessing quality, and end-to-end evaluation necessitates careful orchestration of numerous components, which in turn require domain expertise that is not universally available, particularly outside specialized research centers. Evaluation requires atlas registration, as well as longitudinal registration, which are robust to mass effect and postsurgical deformation, as even small misregistrations can meaningfully distort voxel-level overlap metrics, especially for small recurrences. Tumor segmentation is another necessary component of such pipelines and is particularly challenging in the presence of edema, resection cavities, and postoperative changes. High-quality delineations and quality control are difficult to ensure when medical expertise is not readily available. In addition, many growth models require tissue classification of the entire brain to modulate tumor cell migration according to the underlying tissue class. This segmentation is especially challenging in the tumor region, since images of the healthy brain before disease onset are typically unavailable. By making the full end-to-end pipeline open source, we aim to lower these barriers and make rigorous comparative evaluation more approachable and reproducible for the broader community.

% Paragraph 4 (Key challenge - evaluation/metrics)
% In general, higher specificity without iso-volumetry does not mean less background covered
A further challenge for benchmarking recurrence prediction in a radiotherapy context is the choice of evaluation metrics. In our setting, the percentage-based definition of recurrence coverage is equivalent to the sensitivity, since the true positives correspond to the overlap between the target volume and recurrence, and the sum of true positives and false negatives equals the recurrence volume. However, a framework based solely on sensitivity and specificity is insufficient. Because the class imbalance between normal tissue and tumor is high, simply increasing the predicted target volume can substantially boost sensitivity while only marginally reducing specificity. We therefore require target volumes to be iso-volumetric, forcing models to redistribute a fixed treatment volume toward regions that increase both sensitivity and specificity, rather than benefiting from trivial volume inflation. At the same time, this geometric evaluation does not capture additional factors relevant to radiotherapy planning, such as dose falloff and organ-at-risk constraints, which will be important to incorporate in future dose-aware studies.

% Paragraph 5 (Limitations)
% (from PINN paper) While registration works well in many cases, some errors may arise near the ventricles, especially when there is significant deformation of the brain due to the tumor. For a comparison of the results using affine and diffeomorphic registration, please see Appendix B. It remains an active area of research to design registration methods that robustly account for significant distortions due to tumor mass effect and post-surgical resection (Lipková et al., 2019).
Several limitations of this study should be acknowledged. Although our cohort size is large relative to prior studies, it remains modest compared with typical deep learning requirements, particularly under multi-center domain shift and heterogeneous acquisition protocols. Continued data collection and responsible data sharing by medical institutions are essential to enable larger, more representative cohorts and more definitive comparative evaluations. Additionally, the current benchmark is constrained to routinely acquired standard MRI (with diffusion only available for a subset), and does not yet incorporate modalities such as PET, perfusion, or MR spectroscopy, which may provide complementary biological information and have been reported to improve model predictions in related work~\cite{balcerak2024physics, weidner2024spatial, tran2025novel}. Although integration of additional modalities into PREDICT-GBM is straightforward, there is currently insufficient longitudinal data including these modalities. Evaluation is also limited by the quality of the longitudinal registration, mapping the recurrence to preoperative space. Our framework uses the 2022 MICCAI BraTSReg challenge winning algorithm, which achieved a median absolute error of $1.64\,\mathrm{mm}$~\cite{mok2022robust}. The benchmark also evaluates geometric target definitions rather than delivered dose distributions. Therefore, the observed improvement in recurrence coverage cannot be directly interpreted as expected gains in survival, reduced toxicity, or improved quality of life. Ultimately, clinical impact will require validation and integration into treatment planning workflows, which could include strategies for dose painting, boosting, or other dose-aware adaptations within model-predicted high-risk regions.
Moreover, the follow-up MRIs used for evaluation reflect outcomes under standard-of-care treatment. Consequently, observed recurrence patterns can be interpreted as a spatial proxy for regions of elevated residual tumor burden. In this context, predicting the recurrence core corresponds to identifying high-risk areas that were insufficiently treated by uniform margin–based irradiation, providing a clinically meaningful target. However, this retrospective evaluation should not be conflated with prospective assessment of fully model-based radiation plans, where altered dose distributions may lead to different recurrence patterns and would require dedicated clinical validation.
Finally, model validation is limited by the lack of ground truth in non-MRI-visible tumor regions. This could be addressed by incorporating spatially resolved biopsy data obtained during surgery, enabling calibration of predicted cell density maps beyond the imaging-defined lesion.

\begin{figure*}[t]
\centering
\includegraphics[width=.99\linewidth, trim = 1.5cm 13.0cm 1.0cm 4.3cm, clip]{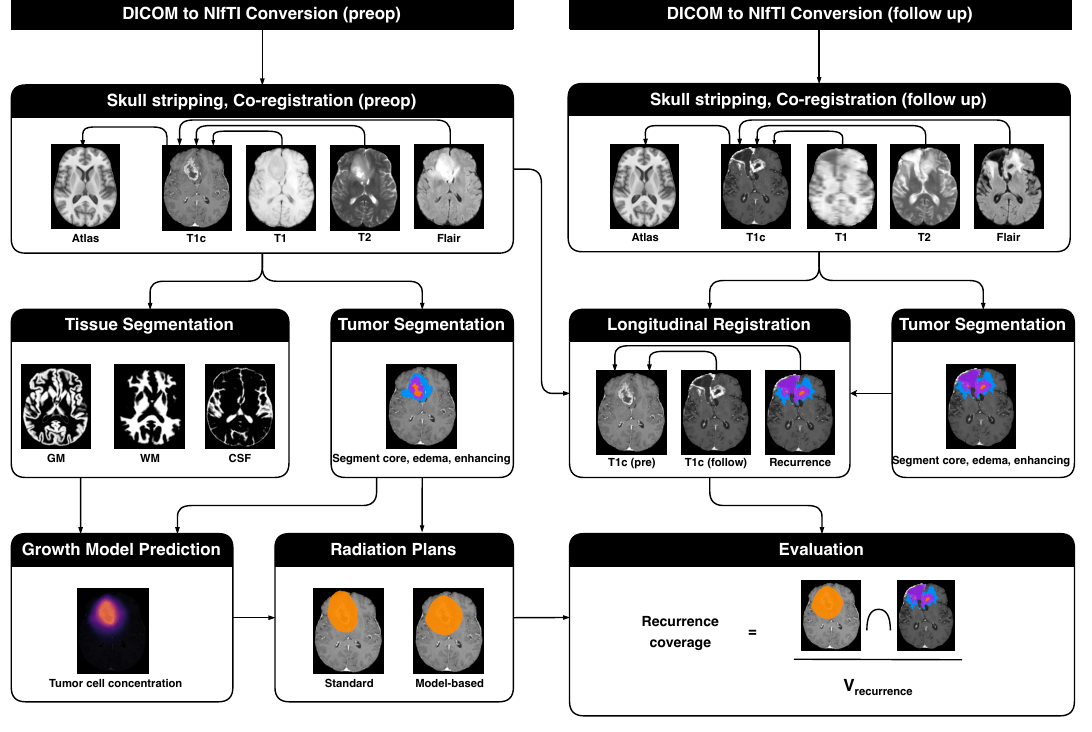}
\caption{Visualization of the PREDICT-GBM pipeline. After converting DICOM inputs to NIfTI format, the skull is stripped, and the modalities are co-registered to the SRI24 atlas space for both the pre-operative and follow-up exams. The tumor is then segmented into enhancing, necrotic recurrence, and peritumoral edema for both exams. For the pre-operative exam, the brain tissue is further segmented into white matter, gray matter, and cerebrospinal fluid. The preoperative tumor and tissue segmentations are then used as input for the models. The follow-up segmentation is deformably registered to the pre-operative exam. Lastly, radiation plans (incl. \ac{soc}) are generated, and their coverage is computed on the registered recurrence.}
\label{fig:pipeline}
\end{figure*}

% Paragraph 6 (Future works)
Looking forward, PREDICT-GBM is designed to accelerate model development and translation by enabling standardized, reproducible evaluation on any suitable cohort. We release anonymized longitudinal MRI data for 142 patients, thereby increasing the availability of public datasets suitable for growth modeling and recurrence prediction research. We hope that in the future, more medical institutions will consider releasing their data, as this is essential for enabling translational research. Beyond increasing MRI cohort sizes, integrating spatially resolved biopsy data could be invaluable for calibrating growth models and for more precisely defining tumors and recurrences within and beyond the MRI-visible region. Methodologically, future growth models could incorporate biophysical mechanisms beyond diffusion–reaction frameworks. For instance, integrating mechanistic mass effect could yield better estimates of tumor cell concentration, particularly for large tumors. Multi-compartment models that distinguish among tumor cell types, such as hypoxic, proliferative, or invasive, could better capture the heterogeneous nature of glioblastoma. In addition, extending physics-informed models to explicitly account for multifocal disease, via multiple tumor seeds or more flexible initialization strategies that allow spatially separated baseline tumor regions, could improve their applicability to patients with multifocal tumors. Our platform naturally supports extending preprocessing to additional modalities suitable for such models, such as diffusion tensor imaging, perfusion, or PET. Diffusion imaging could yield information about local, anisotropic diffusivity beyond the typical fixed values assumed for gray and white brain matter, while perfusion imaging provides information about oxygenation. PET, which indicates regions of metabolic activity, in particular, has been shown to significantly benefit growth models, although available data is scarce~\cite{balcerak2024physics, weidner2024spatial}. In general, growth models scale with the number of modalities, as each provides additional information about the underlying biophysical processes. Highly data-driven models, such as U-Nets, on the other hand, scale with the number of subjects, underlining the need for further data sharing. Finally, future evaluations should move beyond geometric targets toward dose-aware planning: model-derived cell density or risk maps could guide dose painting or boosting, delivering higher doses to regions of high predicted tumor cell concentration while respecting organ-at-risk constraints and realistic dose falloff.

% Paragraph 7 (Conclusion)
PREDICT-GBM provides a standardized, open-source framework that advances the state of the art in personalized glioblastoma radiotherapy planning and drives the development of new models, as demonstrated by our U-Net. By unifying data processing and providing reproducible benchmarks, PREDICT-GBM removes critical barriers to entry for researchers. Our results show that both physics-informed and data-driven models can yield radiotherapy target volumes with significantly higher coverage of regions of later recurrence than the current SOC. With our framework, we hope to foster research in both directions, ultimately creating the necessary evidence base to transition computational growth models from experimental tools into valid, personalized clinical solutions.

% ----------------------------------------------------------------------------------------------------
% Methods
% ----------------------------------------------------------------------------------------------------
\section{Methods}
\label{sec:methods}

\paragraph{Pipeline}
Evaluating tumor growth models for radiation planning requires numerous processing steps. A schematic overview of the PREDICT-GBM pipeline is shown in Figure~\ref{fig:pipeline}. Tools and dependencies were chosen based on performance and runtime considerations, with a preference for open-source solutions. Likewise, our entire pipeline is open source, providing researchers with a plug-and-play framework for growth model evaluation\footnote{\href{www.github.com/BrainLesion/PredictGBM}{github.com/BrainLesion/PredictGBM}}. In the following, we describe the different modules of our pipeline in the order of execution.

\paragraph{Inclusion criteria}
Subjects were curated from three sources: \mbox{LUMIERE}~\cite{suter2022lumiere}, RHUH~\cite{cepeda2023rio}, and TUM-GBM~\cite{weidner2024spatial, balcerak2025individualizing}. The inclusion criteria required (i) a pre-treatment MRI exam, (ii) at least one follow-up exam demonstrating early post-treatment recurrence, (iii) availability of standard MRI sequences for each exam (\ac{t1}, \ac{t1c}, \ac{t2}, and \ac{flair}). Follow-up exams were required to be acquired at least 12 weeks after the treatment baseline exam, as pseudo-progression is most commonly observed in these first weeks of treatment~\cite{wen2023rano}. For patients with multiple follow-up exams, we identified the exam with the best resolution that met the other criteria for inclusion. To minimize site-specific preprocessing variability, we preferred the least-processed image format available (typically raw DICOM). All listed datasets were reviewed by an expert to confirm the presence of visible recurrence in the follow-up MRI scans.

\paragraph{Image preprocessing}
Images were converted from DICOM to NIfTI using dcm2niix~\cite{Li2016}. All modalities were normalized and co-registered to the anatomical SRI24 atlas space~\cite{rohlfing2010sri24} using the BrainLes preprocessing module~\cite{kofler2025brainlesion}. Within each exam, modalities were first rigidly registered to the \ac{t1c} image and then co-registered to atlas space using mutual information as metric and the ANTs library as backend~\cite{avants2009advanced}. Skull stripping was performed with HD-BET~\cite{isensee2019automated}. For longitudinal alignment, the follow-up \ac{t1c} was deformably registered to the pre-treatment \ac{t1c} using an implementation of the winning algorithm of the 2022 MICCAI BraTSReg challenge. This method uses a
conditional deep Laplacian pyramid image registration network with a forward-backward consistency constraint, followed by a non-linear, multi-scale instance optimization using a normalized cross-correlation loss~\cite{mok2022robust}. For comparison (see Figure~\ref{fig:longitudinal_registration}), we also implemented a SyN registration using ANTs \ac{syn}~\cite{avants2008symmetric}, with optimal settings identified in the BraTSReg challenge~\cite{baheti2021brain}, as well as an alternative instance optimization 
minimizing a mean squared intensity error with diffusion-style smoothness penalty. The resulting transforms were applied consistently to all follow-up modalities and derived segmentations to ensure voxel-wise correspondence in the pre-treatment space. For patients with \ac{dti} as only diffusion modality, \ac{adc} was derived via \ac{md}.

\paragraph{Tumor segmentation}
Tumor subregion segmentations were obtained from the preprocessed MRI using the \ac{brats} pipeline~\cite{kofler2025brats}. Pre-treatment segmentation used the \ac{brats} 2023 winning method, an ensemble trained with heavy data augmentation~\cite {ferreira2024we}. Follow-up segmentation used the \ac{brats} 2024 winning method for post-treatment glioma MRI~\cite{de20242024}. Segmentations included necrosis, contrast-enhancing tumor, and peritumoral edema, enabling evaluation of recurrence under multiple clinically relevant definitions. The presence of a tumor was verified by a medical expert for all cases.

\paragraph{Brain tissue segmentation}
Several models require brain tissue segmentations to inform biophysical parameters, such as local diffusivity, which, in turn, affect cell migration. Because tumor and treatment effects disrupt local anatomy, we used an atlas-based approach to obtain tissue priors in patient space. Specifically, the SRI24 atlas was deformably registered to each subject using ANTs \ac{syn}~\cite{avants2008symmetric}, and atlas tissue probability maps were warped into patient space using the obtained transformation. Tissue labels (gray matter, white matter, \ac{csf}) were derived from the warped probability maps and were used as model inputs where required.

\paragraph{Models}
% Could expand on inputs of individual models
We evaluated a diverse set of approaches spanning biophysical reaction-diffusion growth modeling and purely data-driven recurrence prediction. Models that estimate a tumor cell concentration map in our benchmark are based on Fisher-KPP reaction-diffusion PDEs (GliODIL, \ac{loti}, PINN-GBM, \ac{lmi}), while the second family of models directly predicts regions of subsequent recurrence rather than an underlying tumor cell concentration (GlioMap, U-Net, nnU-Net):

\begin{itemize}

\item GliODIL is a glioblastoma growth model that predicts a spatiotemporal distribution for the underlying tumor cell concentration. It combines a data loss term with a reaction-diffusion physics term obtained by discretizing the Fisher-KPP PDE according to the \ac{odil} framework to leverage gradient descent and automatic differentiation with deep-learning optimizers~\cite{balcerak2025individualizing, karnakov2022optimizing}. In this way, both data and physics-based constraints are softly assimilated into the solution.

\item Lightweight Optimization for Estimating Tumor Infiltration (LOTI) estimates a tumor cell concentration by fitting it to the MRI-derived tumor segmentation while enforcing a smooth solution. It uses the Adam optimizer to minimize a combined physics and data loss term, achieving fast runtimes. While the data loss drives the solution to match the MRI-visible tumor, the Dirichlet Energy loss term penalizes large spatial gradients, encouraging physically plausible, diffusion-like solutions as expected from reaction-diffusion models~\cite{weidner2024spatial}.

\item PINN-GBM uses a \ac{pinn}~\cite{raissi2019physics} to solve the inverse Fisher-KPP problem and infer patient-specific parameters and the tumor cell concentration from MRI images. First, patient-specific growth parameters are estimated via grid search, and the Fisher–KPP PDE is solved with these parameters using a finite difference method (FDM) together with the diffuse domain method to handle boundary conditions~\cite{li2009solving}. The solution is then used to pre-train a fully connected neural network that represents the tumor cell concentration in the output layer. Finally, the network is fine-tuned on the segmentation data with trainable patient-specific parameters in the loss function, yielding the final tumor cell concentration and growth parameters~\cite{zhang2025personalized}.

\item Learn-Morph-Infer (LMI) trains a convolutional neural network on simulated data to learn the inverse mapping from MRI to growth parameters in a PDE, such as diffusion coefficient and proliferation rate. The parameters are then used to generate a tumor cell concentration via a conventional PDE solver. The framework also supports more complex PDEs considering mass effect beyond the typical reaction diffusion equations~\cite{ezhov2023learn}.

\item GlioMap is a voxel-wise recurrence prediction model focused on the peritumoral brain region. It extracts voxel-wise radiomic features from MRI modalities, including \ac{adc}, to estimate the recurrence probability for each voxel using ML classifiers~\cite{cepeda2023predicting, cepeda2023evaluation, cepeda2023machine, cepeda2025radiomics}. Training was performed using undersampling to address the strong class imbalance between peritumoral and recurrence voxels, and a distance-based correction factor was applied to penalize the predicted probabilities for voxels far from the resection cavity.

\item nnU-Net is a self-configuring framework that was used as a baseline~\cite{isensee2021nnu}. The baseline nnU-Net was trained using 3-fold cross-validation on our combined dataset, yielding 169 training and 84 test cases per nnU-Net training fold. The data splits were stratified according to the original datasets to account for domain shift across centers. Inputs were the pre-operative tumor segmentation and tissue maps.

\end{itemize}

The PREDICT-GBM repository contains containerized implementations and instructions for including new models, enabling straightforward extension. Table~\ref{tab:growth_models} summarizes runtime and memory requirements. Although GliODIL has significantly longer runtimes, it is also the only model that directly yields a temporal distribution.

% Moved U-Net to its own paragraph and included more details, removed "cross-validation" terms
\paragraph{Custom U-Net}
Our custom U-Net is trained using a deep-ensemble strategy with three ensemble members, each using different random initialization and randomized data shuffling~\cite{ronneberger2015u, lakshminarayanan_simple_2017}. For each ensemble member, we trained five U-Nets on distinct train/validation/test splits, resulting in a total of 15 trained U-Nets. To obtain unbiased out-of-sample predictions for all patients, each prediction is generated from the model for which the corresponding patient belongs to the held-out test set, ensuring that all results are obtained on unseen data. We used an extended dataset of 327 patients, yielding 197 training, 65 validation, and 65 test cases per split, stratified by dataset origin. The extended training dataset is supplemented with 84 additional patients from the \ac{cptac}~\cite{CPTAC_GBM_2018}, \ac{ivygap}~\cite{IvyGAP_2016}, \ac{tcgagbm}~\cite{TCGA_GBM_2016}, and \ac{upenngbm}~\cite{UPENN_GBM_2021}. The additional datasets (CPTAC-GBM, IVYGAP, TCGA-GBM, and UPENN-GBM) are not included in our public data release due to licensing restrictions. We used the following inputs: preoperative tumor segmentation, tissue segmentations, preoperative \ac{t1c}, \ac{flair}, and \ac{adc} when available. Channel dropout was used to minimize the impact of missing inputs, along with numerous data augmentation strategies. Random Gaussian edge softening \cite{pornvoraphat2023real} was used to reduce the impact of label noise, transforming the binary segmentations into continuous probability maps. The U-Net was trained using a modified composite loss function inspired by \cite{tran2025novel}, combining a \ac{bce} loss with an individualized \ac{pcc} loss derived from the Tversky loss. The \ac{pcc} loss dynamically adjusts the loss hyperparameters for each patient based on tumor size to address class imbalance. For further details on the implementation, we refer to the supplementary material.

% Could add formula for loss but it would be the same as in J. Lupo

\begin{table}[t]
\centering
\begin{tabular}{@{\hspace{.1cm}}l@{\hspace{.5cm}}l@{\hspace{.5cm}}c@{\hspace{.5cm}}l@{\hspace{.1cm}}}
\toprule
\textbf{Model}    & \textbf{Inference}      & \textbf{Training}      & \textbf{Memory} \\ \midrule
LOTI              & $\sim1$ min             & -                      & $\sim2$ GB      \\
GliODIL           & $\sim50$ min            & -                      & $\sim18$ GB     \\
U-Net             & $<1$ s                  & $\sim 50$ h            & $\sim1$ GB      \\
PINN-GBM          & $\sim20$ min            & -                      & $\sim20$ GB     \\
LMI               & $\sim5$ min             & $\sim130$ h            & $\sim1$ GB      \\
nnU-Net           & $\sim5$ s               & $\sim50$ h             & $\sim2$ GB      \\ \bottomrule
\end{tabular}
\vspace{.4cm}
\caption{Inference time per case, total training time, and memory requirements during inference for different models. Inference and training times are reported on a Quadro~RTX~8000 GPU.}
\label{tab:growth_models}
\end{table}

\paragraph{Radiation plans}
We compared model-derived target volumes against a margin-based \ac{soc} \ac{rt} target definition. The \ac{soc} target volume was generated by isotropically dilating the segmented tumor core by 15~mm within the brain mask, consistent with contemporary guideline-based \ac{ctv} expansions~\cite{Niyazi2023}. For models producing continuous voxel-wise scores, either tumor cell concentration maps (GliODIL, \ac{loti}, PINN-GBM, \ac{lmi}) or recurrence probability maps (GlioMap, U-Net, nnU-Net), we constructed iso-volumetric targets by selecting the top-$k$ highest-scoring voxels such that the resulting target volume matched the \ac{soc} volume. This iso-volumetric constraint prevents trivial performance gains via volume inflation and forces models to redistribute a fixed treatment volume toward higher-risk regions~\cite{buti2024influence}. For GlioMap, which provides scores only in the peritumoral region, we added the tumor core to the GlioMap score map prior to top-$k$ selection. When a model output contained fewer non-zero (or positive) voxels than required to reach the \ac{soc} volume, we expanded the prediction isotropically by binarizing the output, computing a distance transform, and adding voxels in order of increasing distance to the predicted region until iso-volumetry was achieved.

\paragraph{Recurrence definitions and evaluation}
Evaluation was performed by computing the recurrence coverage in the pre-treatment reference space using the longitudinal transforms described above. We report two recurrence definitions: (i) enhancing recurrence, and (ii) enhancing plus necrotic recurrence (recurrence core). For each definition, recurrence coverage, or equivalently, the sensitivity, was computed as:
\[
\mathrm{Coverage} = \frac{| V_{target} \cap V_{rec}|}{| V_{rec} |} = \frac{\mathrm{TP}}{\mathrm{TP + FN}} = \mathrm{Sensitivity}
\]
where $V_{target}$ is the evaluated target volume (SOC or model-based) and $V_{rec}$ is the corresponding follow-up segmentation.

\paragraph{Statistical analysis}
Because recurrence coverage values are bounded and frequently concentrated near 0 or 1, recurrence coverage cannot be assumed to be normally distributed. Thus, paired Wilcoxon signed-rank tests were applied to compare per-patient recurrence coverage between each model-based target and the \ac{soc} target under each recurrence definition. Statistical significance was assessed at $p<0.05$.

% ----------------------------------------------------------------------------------------------------
% Data Availability
% ----------------------------------------------------------------------------------------------------
\section{Data availability}
To foster reproducibility, we provide the exact list of patients and the modalities identified for each public dataset at \url{github.com/BrainLesion/PredictGBM}. Furthermore, we release skull-stripped MRI exams for 142 subjects from the TUM-GBM dataset, ensuring that all 243 patients are publicly available. By providing this information, anyone with access to the public datasets can reproduce our experiments while protecting patient anonymity. In addition to the MRI exams, we provide the processed tumor segmentations, tissue segmentations, and growth model tumor cell maps as NIfTI files at \url{huggingface.co/datasets/LZimmer/PREDICT-GBM}. The LUMIERE dataset is publicly available at \url{doi.org/10.6084/m9.figshare.c.5904905} and the RHUH dataset is publicly available at \url{cancerimagingarchive.net/collection/rhuh-gbm/}.

% ----------------------------------------------------------------------------------------------------
% Code Availability
% ----------------------------------------------------------------------------------------------------
\section{Code availability}
The complete PREDICT-GBM pipeline is publicly available at \url{github.com/BrainLesion/PredictGBM}. The repository provides the full end-to-end workflow, including containerized implementations of the evaluated growth models and recurrence prediction models. Detailed documentation and integration guidelines are provided to facilitate the inclusion of new models.

% ----------------------------------------------------------------------------------------------------
% Abbreviations
% ----------------------------------------------------------------------------------------------------
\section*{Abbreviations}
\printacronyms[heading=none]

% ----------------------------------------------------------------------------------------------------
% References
% ----------------------------------------------------------------------------------------------------
\bibliography{references}

@article{lakshminarayanan_simple_2017,
  title={Simple and scalable predictive uncertainty estimation using deep ensembles},
  author={Lakshminarayanan, Balaji and Pritzel, Alexander and Blundell, Charles},
  journal={Advances in neural information processing systems},
  volume={30},
  year={2017}
}

@dataset{CPTAC_GBM_2018,
   author       = {{National Cancer Institute Clinical Proteomic Tumor Analysis Consortium (CPTAC)}},
   title        = {The Clinical Proteomic Tumor Analysis Consortium Glioblastoma Multiforme Collection (CPTAC-GBM)},
   year         = {2018},
   version      = {16},
   publisher    = {The Cancer Imaging Archive},
   doi          = {10.7937/K9/TCIA.2018.3RJE41Q1},
   url          = {https://doi.org/10.7937/K9/TCIA.2018.3RJE41Q1}
 }

@dataset{IvyGAP_2016,
   author       = {Shah, N. and Feng, X. and Lankerovich, M. and Puchalski, R. B. and Keogh, B.},
   title        = {Data from Ivy Glioblastoma Atlas Project (IvyGAP)},
   year         = {2016},
   publisher    = {The Cancer Imaging Archive},
   doi          = {10.7937/K9/TCIA.2016.XLwaN6nL},
   url          = {https://doi.org/10.7937/K9/TCIA.2016.XLwaN6nL}
 }

@dataset{TCGA_GBM_2016,
   author       = {Scarpace, L. and Mikkelsen, T. and Cha, S. and Rao, S. and Tekchandani, S. and Gutman, D. and Saltz, J. H. and Erickson, B. J. and Pedano, N. and Flanders, A. E. and Barnholtz-Sloan, J. and Ostrom, Q. and Barboriak, D. and Pierce, L. J.},
   title        = {The Cancer Genome Atlas Glioblastoma Multiforme Collection (TCGA-GBM)},
   year         = {2016},
   version      = {5},
   publisher    = {The Cancer Imaging Archive},
   doi          = {10.7937/K9/TCIA.2016.RNYFUYE9},
   url          = {https://doi.org/10.7937/K9/TCIA.2016.RNYFUYE9}
   }

@dataset{UPENN_GBM_2021,
   author       = {Bakas, S. and Sako, C. and Akbari, H. and Bilello, M. and Sotiras, A. and Shukla, G. and Rudie, J. D. and Flores Santamaria, N. and Fathi Kazerooni, A. and Pati, S. and Rathore, S. and Mamourian, E. and Ha, S. M. and Parker, W. and Doshi, J. and Baid, U. and Bergman, M. and Binder, Z. A. and Verma, R. and ... Davatzikos, C.},
   title        = {Multi-parametric magnetic resonance imaging (mpMRI) scans for de novo Glioblastoma (GBM) patients from the University of Pennsylvania Health System (UPENN-GBM)},
   year         = {2021},
   version      = {2},
   publisher    = {The Cancer Imaging Archive},
   doi          = {10.7937/TCIA.709X-DN49},
   url          = {https://doi.org/10.7937/TCIA.709X-DN49}
 }

@article{ulyanov_instance_2017,
  title={Instance normalization: The missing ingredient for fast stylization},
  author={Ulyanov, Dmitry and Vedaldi, Andrea and Lempitsky, Victor},
  journal={arXiv preprint arXiv:1607.08022},
  year={2016}
}

@article{pornvoraphat2023real,
  title={Real-time gastric intestinal metaplasia diagnosis tailored for bias and noisy-labeled data with multiple endoscopic imaging},
  author={Pornvoraphat, Passin and Tiankanon, Kasenee and Pittayanon, Rapat and Sunthornwetchapong, Phanukorn and Vateekul, Peerapon and Rerknimitr, Rungsun},
  journal={Computers in Biology and Medicine},
  volume={154},
  pages={106582},
  year={2023},
  publisher={Elsevier}
}

@article{stupp2005radiotherapy,
  title={Radiotherapy plus concomitant and adjuvant temozolomide for glioblastoma},
  author={Stupp, Roger and Mason, Warren P and Van Den Bent, Martin J and Weller, Michael and Fisher, Barbara and Taphoorn, Martin JB and Belanger, Karl and Brandes, Alba A and Marosi, Christine and Bogdahn, Ulrich and others},
  journal={New England Journal of Medicine},
  volume={352},
  number={10},
  pages={987--996},
  year={2005},
  publisher={Mass Medical Soc}
}

@article{koshy2012improved,
  title={Improved survival time trends for glioblastoma using the SEER 17 population-based registries},
  author={Koshy, Matthew and Villano, John L and Dolecek, Therese A and Howard, Andrew and Mahmood, Usama and Chmura, Steven J and Weichselbaum, Ralph R and McCarthy, Bridget J},
  journal={Journal of Neuro-Oncology},
  volume={107},
  pages={207--212},
  year={2012},
  publisher={Springer}
}

@article{brown2022survival,
  title={Survival outcomes and prognostic factors in glioblastoma},
  author={Brown, Nicholas F and Ottaviani, Diego and Tazare, John and Gregson, John and Kitchen, Neil and Brandner, Sebastian and Fersht, Naomi and Mulholland, Paul},
  journal={Cancers},
  volume={14},
  number={13},
  pages={3161},
  year={2022},
  publisher={MDPI}
}

@article{mohammed2022survival,
  title={Survival and quality of life analysis in glioblastoma multiforme with adjuvant chemoradiotherapy: a retrospective study},
  author={Mohammed, Soniya and Dinesan, M and Ajayakumar, T},
  journal={Reports of Practical Oncology and Radiotherapy},
  volume={27},
  number={6},
  pages={1026--1036},
  year={2022}
}

@article{ostrom2018females,
  title={Females have the survival advantage in glioblastoma},
  author={Ostrom, Quinn T and Rubin, Joshua B and Lathia, Justin D and Berens, Michael E and Barnholtz-Sloan, Jill S},
  journal={Neuro-oncology},
  volume={20},
  number={4},
  pages={576--577},
  year={2018},
  publisher={Oxford University Press US}
}

@article{grochans2022epidemiology,
  title={Epidemiology of glioblastoma multiforme--literature review},
  author={Grochans, Szymon and Cybulska, Anna Maria and Simi{\'n}ska, Donata and Korbecki, Jan and Kojder, Klaudyna and Chlubek, Dariusz and Baranowska-Bosiacka, Irena},
  journal={Cancers},
  volume={14},
  number={10},
  pages={2412},
  year={2022},
  publisher={MDPI}
}

@article{weller2013standards,
  title={Standards of care for treatment of recurrent glioblastoma—are we there yet?},
  author={Weller, Michael and Cloughesy, Timothy and Perry, James R and Wick, Wolfgang},
  journal={Neuro-oncology},
  volume={15},
  number={1},
  pages={4--27},
  year={2013},
  publisher={Oxford University Press}
}

@article{metz2023towards,
  title={Towards image-based personalization of glioblastoma therapy a clinical and biological validation study of a novel, deep learning-driven tumor growth model},
  author={Metz, Marie-Christin and Ezhov, Ivan and Zimmer, Lucas and Peeken, Jan C and Buchner, Josef A and Lipkova, Jana and Kofler, Florian and Waldmannstetter, Diana and Delbridge, Claire and Diehl, Christian and others},
  year={2023}
}

@inproceedings{menze2011image,
  title={A generative approach for image-based modeling of tumor growth},
  author={Menze, Bjoern H and Van Leemput, Koen and Honkela, Antti and Konukoglu, Ender and Weber, Marc-Andr{\'e} and Ayache, Nicholas and Golland, Polina},
  booktitle={Biennial International Conference on Information Processing in Medical Imaging},
  pages={735--747},
  year={2011},
  organization={Springer}
}

@article{konukoglu2009image,
  title={Image guided personalization of reaction-diffusion type tumor growth models using modified anisotropic eikonal equations},
  author={Konukoglu, Ender and Clatz, Olivier and Menze, Bjoern H and Stieltjes, Bram and Weber, Marc-Andr{\'e} and Mandonnet, Emmanuel and Delingette, Herv{\'e} and Ayache, Nicholas},
  journal={IEEE Transactions on Medical Imaging},
  volume={29},
  number={1},
  pages={77--95},
  year={2009},
  publisher={IEEE}
}

@article{frieboes2007computer,
  title={Computer simulation of glioma growth and morphology},
  author={Frieboes, Hermann B and Lowengrub, John S and Wise, S and Zheng, X and Macklin, Paul and Bearer, Elaine L and Cristini, Vittorio},
  journal={Neuroimage},
  volume={37},
  pages={S59--S70},
  year={2007},
  publisher={Elsevier}
}

@article{lipkova2019personalized,
  title={Personalized radiotherapy design for glioblastoma: integrating mathematical tumor models, multimodal scans, and Bayesian inference},
  author={Lipkov{\'a}, Jana and Angelikopoulos, Panagiotis and Wu, Stephen and Alberts, Esther and Wiestler, Benedikt and Diehl, Christian and Preibisch, Christine and Pyka, Thomas and Combs, Stephanie E and Hadjidoukas, Panagiotis and others},
  journal={IEEE Transactions on Medical Imaging},
  volume={38},
  number={8},
  pages={1875--1884},
  year={2019},
  publisher={IEEE}
}

@article{mang2012biophysical,
  title={Biophysical modeling of brain tumor progression: From unconditionally stable explicit time integration to an inverse problem with parabolic PDE constraints for model calibration},
  author={Mang, Andreas and Toma, Alina and Schuetz, Tina A and Becker, Stefan and Eckey, Thomas and Mohr, Christian and Petersen, Dirk and Buzug, Thorsten M},
  journal={Medical Physics},
  volume={39},
  number={7Part1},
  pages={4444--4459},
  year={2012},
  publisher={Wiley Online Library}
}

@article{scheufele2020image,
  title={Image-driven biophysical tumor growth model calibration},
  author={Scheufele, Klaudius and Subramanian, Shashank and Mang, Andreas and Biros, George and Mehl, Miriam},
  journal={SIAM journal on scientific computing: a publication of the Society for Industrial and Applied Mathematics},
  volume={42},
  number={3},
  pages={B549},
  year={2020},
}

@article{scheufele2020fully,
  title={Fully automatic calibration of tumor-growth models using a single mpMRI scan},
  author={Scheufele, Klaudius and Subramanian, Shashank and Biros, George},
  journal={IEEE Transactions on Medical Imaging},
  volume={40},
  number={1},
  pages={193--204},
  year={2020},
  publisher={IEEE}
}

@inproceedings{subramanian2020multiatlas,
  title={Multiatlas calibration of biophysical brain tumor growth models with mass effect},
  author={Subramanian, Shashank and Scheufele, Klaudius and Himthani, Naveen and Biros, George},
  booktitle={Medical Image Computing and Computer Assisted Intervention--MICCAI 2020: 23rd International Conference, Lima, Peru, October 4--8, 2020, Proceedings, Part II 23},
  pages={551--560},
  year={2020},
  organization={Springer}
}

@article{subramanian2022ensemble,
  title={Ensemble inversion for brain tumor growth models with mass effect},
  author={Subramanian, Shashank and Ghafouri, Ali and Scheufele, Klaudius Matthias and Himthani, Naveen and Davatzikos, Christos and Biros, George},
  journal={IEEE Transactions on Medical Imaging},
  volume={42},
  number={4},
  pages={982--995},
  year={2022},
  publisher={IEEE}
}

@article{unkelbach2014radiotherapy,
  title={Radiotherapy planning for glioblastoma based on a tumor growth model: improving target volume delineation},
  author={Unkelbach, Jan and Menze, Bjoern H and Konukoglu, Ender and Dittmann, Florian and Le, Matthieu and Ayache, Nicholas and Shih, Helen A},
  journal={Physics in Medicine \& Biology},
  volume={59},
  number={3},
  pages={747},
  year={2014},
  publisher={IOP Publishing}
}

@article{ezhov2021geometry,
  title={Geometry-aware neural solver for fast Bayesian calibration of brain tumor models},
  author={Ezhov, Ivan and Mot, Tudor and Shit, Suprosanna and Lipkova, Jana and Paetzold, Johannes C and Kofler, Florian and Pellegrini, Chantal and Kollovieh, Marcel and Navarro, Fernando and Li, Hongwei and others},
  journal={IEEE Transactions on Medical Imaging},
  volume={41},
  number={5},
  pages={1269--1278},
  year={2021},
  publisher={IEEE}
}

@article{viguerie2022data,
  title={Data-driven simulation of Fisher--Kolmogorov tumor growth models using dynamic mode decomposition},
  author={Viguerie, Alex and Grave, Mal{\'u} and Barros, Gabriel F and Lorenzo, Guillermo and Reali, Alessandro and Coutinho, Alvaro LGA},
  journal={Journal of Biomechanical Engineering},
  volume={144},
  number={12},
  pages={121001},
  year={2022},
  publisher={American Society of Mechanical Engineers}
}

@article{martens2022deep,
  title={Deep learning for reaction-diffusion glioma growth modeling: Towards a fully personalized model?},
  author={Martens, Corentin and Rovai, Antonin and Bonatto, Daniele and Metens, Thierry and Debeir, Olivier and Decaestecker, Christine and Goldman, Serge and Van Simaeys, Gaetan},
  journal={Cancers},
  volume={14},
  number={10},
  pages={2530},
  year={2022},
  publisher={MDPI}
}

@inproceedings{ezhov2019neural,
  title={Neural parameters estimation for brain tumor growth modeling},
  author={Ezhov, Ivan and Lipkova, Jana and Shit, Suprosanna and Kofler, Florian and Collomb, Nore and Lemasson, Benjamin and Barbier, Emmanuel and Menze, Bjoern},
  booktitle={Medical Image Computing and Computer Assisted Intervention--MICCAI 2019: 22nd International Conference, Shenzhen, China, October 13--17, 2019, Proceedings, Part II 22},
  pages={787--795},
  year={2019},
  organization={Springer}
}

@inproceedings{pati2021estimating,
  title={Estimating glioblastoma biophysical growth parameters using deep learning regression},
  author={Pati, Sarthak and Sharma, Vaibhav and Aslam, Heena and Thakur, Siddhesh P and Akbari, Hamed and Mang, Andreas and Subramanian, Shashank and Biros, George and Davatzikos, Christos and Bakas, Spyridon},
  booktitle={Brainlesion: Glioma, Multiple Sclerosis, Stroke and Traumatic Brain Injuries: 6th International Workshop, BrainLes 2020, Held in Conjunction with MICCAI 2020, Lima, Peru, October 4, 2020, Revised Selected Papers, Part I 6},
  pages={157--167},
  year={2021},
  organization={Springer}
}

@article{ezhov2023learn,
  title={Learn-Morph-Infer: a new way of solving the inverse problem for brain tumor modeling},
  author={Ezhov, Ivan and Scibilia, Kevin and Franitza, Katharina and Steinbauer, Felix and Shit, Suprosanna and Zimmer, Lucas and Lipkova, Jana and Kofler, Florian and Paetzold, Johannes C and Canalini, Luca and others},
  journal={Medical Image Analysis},
  volume={83},
  pages={102672},
  year={2023},
  publisher={Elsevier}
}

@article{weidner2024learnable,
  title={A learnable prior improves inverse tumor growth modeling},
  author={Weidner, Jonas and Ezhov, Ivan and Balcerak, Michal and Metz, Marie-Christin and Litvinov, Sergey and Kaltenbach, Sebastian and Feiner, Leonhard and Lux, Laurin and Kofler, Florian and Lipkova, Jana and others},
  journal={IEEE Transactions on Medical Imaging},
  year={2024},
  publisher={IEEE}
}

@inproceedings{weidner2024rapid,
  title={Rapid Personalization of PDE-Based Tumor Growth using a Differentiable Forward Model},
  author={Weidner, Jonas and Ezhov, Ivan and Balcerak, Michal and Wiestler, Benedikt and others},
  booktitle={Medical Imaging with Deep Learning},
  year={2024}
}

@article{zhang2025personalized,
  title={Personalized predictions of Glioblastoma infiltration: Mathematical models, Physics-Informed Neural Networks and multimodal scans},
  author={Zhang, Ray Zirui and Ezhov, Ivan and Balcerak, Michal and Zhu, Andy and Wiestler, Benedikt and Menze, Bjoern and Lowengrub, John S},
  journal={Medical Image Analysis},
  volume={101},
  pages={103423},
  year={2025},
  publisher={Elsevier}
}

@inproceedings{weidner2024spatial,
  title={A Lightweight Optimization Framework for Estimating 3D Brain Tumor Infiltration},
  author={Weidner, Jonas and Balcerak, Michal and Ezhov, Ivan and Datchev, Andr{\'e} and Lux, Laurin and Zimmer, Lucas and Rueckert, Daniel and Menze, Bj{\"o}rn and Wiestler, Benedikt},
  booktitle={International Workshop on Computational Mathematics Modeling in Cancer Analysis},
  pages={1--10},
  year={2025},
  organization={Springer}
}

@article{menze2014multimodal,
  title     = {The multimodal brain tumor image segmentation benchmark (BRATS)},
  author    = {Menze, Bjoern H and Jakab, Andras and Bauer, Stefan and Kalpathy-Cramer, Jayashree and Farahani, Keyvan and Kirby, Justin and Burren, Yuliya and Porz, Nicole and Slotboom, Johannes and Wiest, Roland and others},
  journal   = {IEEE Transactions on Medical Imaging},
  volume    = {34},
  number    = {10},
  pages     = {1993--2024},
  year      = {2014},
  publisher = {IEEE},
}

@article{kofler2020brats,
  title={Brats toolkit: translating brats brain tumor segmentation algorithms into clinical and scientific practice},
  author={Kofler, Florian and Berger, Christoph and Waldmannstetter, Diana and Lipkova, Jana and Ezhov, Ivan and Tetteh, Giles and Kirschke, Jan and Zimmer, Claus and Wiestler, Benedikt and Menze, Bjoern H},
  journal={Frontiers in neuroscience},
  volume={14},
  pages={125},
  year={2020},
  publisher={Frontiers Media SA}
}

@article{kofler2025brats,
  title={BraTS orchestrator: Democratizing and Disseminating state-of-the-art brain tumor image analysis},
  author={Kofler, Florian and Rosier, Marcel and Astaraki, Mehdi and Baid, Ujjwal and M{\"o}ller, Hendrik and Buchner, Josef A and Steinbauer, Felix and Oswald, Eva and de la Rosa, Ezequiel and Ezhov, Ivan and others},
  journal={arXiv preprint arXiv:2506.13807},
  year={2025}
}

@article{fischl2012freesurfer,
  title={FreeSurfer},
  author={Fischl, Bruce},
  journal={Neuroimage},
  volume={62},
  number={2},
  pages={774--781},
  year={2012},
  publisher={Elsevier}
}

@article{billot2023synthseg,
  title={SynthSeg: Segmentation of brain MRI scans of any contrast and resolution without retraining},
  author={Billot, Benjamin and Greve, Douglas N and Puonti, Oula and Thielscher, Axel and Van Leemput, Koen and Fischl, Bruce and Dalca, Adrian V and Iglesias, Juan Eugenio and others},
  journal={Medical image analysis},
  volume={86},
  pages={102789},
  year={2023},
  publisher={Elsevier}
}

@article{henschel2020fastsurfer,
  title={Fastsurfer-a fast and accurate deep learning based neuroimaging pipeline},
  author={Henschel, Leonie and Conjeti, Sailesh and Estrada, Santiago and Diers, Kersten and Fischl, Bruce and Reuter, Martin},
  journal={NeuroImage},
  volume={219},
  pages={117012},
  year={2020},
  publisher={Elsevier}
}

@article{zhang2001segmentation,
  title={Segmentation of brain MR images through a hidden Markov random field model and the expectation-maximization algorithm},
  author={Zhang, Yongyue and Brady, Michael and Smith, Stephen},
  journal={IEEE Transactions on Medical Imaging},
  volume={20},
  number={1},
  pages={45--57},
  year={2001},
  publisher={Ieee}
}

@article{tustison_antsx_2021,
	title = {The {ANTsX} ecosystem for quantitative biological and medical imaging},
	volume = {11},
	issn = {2045-2322},
	url = {https://doi.org/10.1038/s41598-021-87564-6},
	doi = {10.1038/s41598-021-87564-6},
	number = {1},
	journal = {Scientific Reports},
	author = {Tustison, Nicholas J. and Cook, Philip A. and Holbrook, Andrew J. and Johnson, Hans J. and Muschelli, John and Devenyi, Gabriel A. and Duda, Jeffrey T. and Das, Sandhitsu R. and Cullen, Nicholas C. and Gillen, Daniel L. and Yassa, Michael A. and Stone, James R. and Gee, James C. and Avants, Brian B.},
	month = apr,
	year = {2021},
	pages = {9068},
}

@article{cepeda2023rio,
  title={The R{\'\i}o Hortega University Hospital Glioblastoma dataset: A comprehensive collection of preoperative, early postoperative and recurrence MRI scans (RHUH-GBM)},
  author={Cepeda, Santiago and Garc{\'\i}a-Garc{\'\i}a, Sergio and Arrese, Ignacio and Herrero, Francisco and Escudero, Trinidad and Zamora, Tom{\'a}s and Sarabia, Rosario},
  journal={Data in Brief},
  volume={50},
  pages={109617},
  year={2023},
  publisher={Elsevier}
}

@article{suter2022lumiere,
  title={The LUMIERE dataset: Longitudinal Glioblastoma MRI with expert RANO evaluation},
  author={Suter, Yannick and Knecht, Urspeter and Valenzuela, Waldo and Notter, Michelle and Hewer, Ekkehard and Schucht, Philippe and Wiest, Roland and Reyes, Mauricio},
  journal={Scientific data},
  volume={9},
  number={1},
  pages={768},
  year={2022},
  publisher={Nature Publishing Group UK London}
}

@article{Li2016,
  author    = {Xiangrui Li and Paul S. Morgan and John Ashburner and Jolinda Smith and Christopher Rorden},
  title     = {The first step for neuroimaging data analysis: DICOM to NIfTI conversion},
  journal   = {Journal of Neuroscience Methods},
  volume    = {264},
  pages     = {47--56},
  year      = {2016},
  doi       = {10.1016/j.jneumeth.2016.03.001},
  url       = {https://pubmed.ncbi.nlm.nih.gov/26945974/}
}

@article{rohlfing2010sri24,
  title={The SRI24 multichannel atlas of normal adult human brain structure},
  author={Rohlfing, Torsten and Zahr, Natalie M and Sullivan, Edith V and Pfefferbaum, Adolf},
  journal={Human brain mapping},
  volume={31},
  number={5},
  pages={798--819},
  year={2010},
  publisher={Wiley Online Library}
}

@article{avants2009advanced,
  title={Advanced normalization tools (ANTS)},
  author={Avants, Brian B and Tustison, Nick and Song, Gang and others},
  journal={Insight j},
  volume={2},
  number={365},
  pages={1--35},
  year={2009}
}

@article{isensee2019automated,
  title={Automated brain extraction of multisequence MRI using artificial neural networks},
  author={Isensee, Fabian and Schell, Marianne and Pflueger, Irada and Brugnara, Gianluca and Bonekamp, David and Neuberger, Ulf and Wick, Antje and Schlemmer, Heinz-Peter and Heiland, Sabine and Wick, Wolfgang and others},
  journal={Human brain mapping},
  volume={40},
  number={17},
  pages={4952--4964},
  year={2019},
  publisher={Wiley Online Library}
}

@article{avants2008symmetric,
  title={Symmetric diffeomorphic image registration with cross-correlation: evaluating automated labeling of elderly and neurodegenerative brain},
  author={Avants, Brian B and Epstein, Charles L and Grossman, Murray and Gee, James C},
  journal={Medical image analysis},
  volume={12},
  number={1},
  pages={26--41},
  year={2008},
  publisher={Elsevier}
}

@article{ferreira2024we,
  title={How we won brats 2023 adult glioma challenge? just faking it! enhanced synthetic data augmentation and model ensemble for brain tumour segmentation},
  author={Ferreira, Andr{\'e} and Solak, Naida and Li, Jianning and Dammann, Philipp and Kleesiek, Jens and Alves, Victor and Egger, Jan},
  journal={arXiv preprint arXiv:2402.17317},
  year={2024}
}

@inproceedings{ronneberger2015u,
  title={U-net: Convolutional networks for biomedical image segmentation},
  author={Ronneberger, Olaf and Fischer, Philipp and Brox, Thomas},
  booktitle={Medical image computing and computer-assisted intervention--MICCAI 2015: 18th international conference, Munich, Germany, October 5-9, 2015, proceedings, part III 18},
  pages={234--241},
  year={2015},
  organization={Springer}
}

@article{karnakov2022optimizing,
  title={Optimizing a DIscrete Loss (ODIL) to solve forward and inverse problems for partial differential equations using machine learning tools},
  author={Karnakov, Petr and Litvinov, Sergey and Koumoutsakos, Petros},
  journal={arXiv preprint arXiv:2205.04611},
  year={2022}
}

@article{Niyazi2023,
  author = {Niyazi, Maximilian and Andratschke, Nicolaus and Bendszus, Martin and et al.},
  title = {ESTRO–EANO guideline on target delineation and radiotherapy for glioblastoma},
  journal = {Radiotherapy and Oncology},
  volume = {184},
  pages = {109663},
  year = {2023}
}

@article{buti2024influence,
  title={The influence of anisotropy on the clinical target volume of brain tumor patients},
  author={Buti, Gregory and Ajdari, Ali and Hochreuter, Kim and Shih, Helen and Bridge, Christopher P and Sharp, Gregory C and Bortfeld, Thomas},
  journal={Physics in Medicine \& Biology},
  volume={69},
  number={3},
  pages={035006},
  year={2024},
  publisher={IOP Publishing}
}

@article{de20242024,
  title={The 2024 Brain Tumor Segmentation (BraTS) challenge: glioma segmentation on post-treatment MRI},
  author={de Verdier, Maria Correia and Saluja, Rachit and Gagnon, Louis and LaBella, Dominic and Baid, Ujjwall and Tahon, Nourel Hoda and Foltyn-Dumitru, Martha and Zhang, Jikai and Alafif, Maram and Baig, Saif and others},
  journal={arXiv preprint arXiv:2405.18368},
  year={2024}
}

@article{wen2023rano,
  title={RANO 2.0: update to the response assessment in neuro-oncology criteria for high-and low-grade gliomas in adults},
  author={Wen, Patrick Y and Van Den Bent, Martin and Youssef, Gilbert and Cloughesy, Timothy F and Ellingson, Benjamin M and Weller, Michael and Galanis, Evanthia and Barboriak, Daniel P and De Groot, John and Gilbert, Mark R and others},
  journal={Journal of Clinical Oncology},
  volume={41},
  number={33},
  pages={5187--5199},
  year={2023},
  publisher={Wolters Kluwer Health}
}

@article{balcerak2024physics,
  title={Physics-regularized multi-modal image assimilation for brain tumor localization},
  author={Balcerak, Michal and Amiranashvili, Tamaz and Wagner, Andreas and Weidner, Jonas and Karnakov, Petr and Paetzold, Johannes C and Ezhov, Ivan and Koumoutsakos, Petros and Wiestler, Benedikt and others},
  journal={Advances in Neural Information Processing Systems},
  volume={37},
  pages={41909--41933},
  year={2024}
}

@article{balcerak2025individualizing,
  title={Individualizing glioma radiotherapy planning by optimization of a data and physics-informed discrete loss},
  author={Balcerak, Michal and Weidner, Jonas and Karnakov, Petr and Ezhov, Ivan and Litvinov, Sergey and Koumoutsakos, Petros and Amiranashvili, Tamaz and Zhang, Ray Zirui and Lowengrub, John S and Yakushev, Igor and others},
  journal={Nature Communications},
  volume={16},
  number={1},
  pages={5982},
  year={2025},
  publisher={Nature Publishing Group UK London}
}

@article{kofler2025brainlesion,
  title={BrainLesion Suite: A Flexible and User-Friendly Framework for Modular Brain Lesion Image Analysis},
  author={Kofler, Florian and Rosier, Marcel and Astaraki, Mehdi et al.},
  journal={arXiv preprint https://arxiv.org/abs/2507.09036},
  year={2025}
}

@article{Clark2023TCIA,
  title   = {The Cancer Imaging Archive (TCIA): Maintaining and operating a public information repository},
  author  = {Clark, Kenneth W. and Vendt, Bruce A. and Smith, Kirk E. and Freymann, John B. and Kirby, Justin S. and Koppel, Paul and Moore, Stephen M. and Phillips, Stanley R. and Maffitt, David R. and Pringle, Michael and Tarbox, Lawrence and Prior, Fred W.},
  journal = {Journal of Digital Imaging},
  year    = {2013},
  volume  = {26},
  number  = {6},
  pages   = {1045--1057},
  doi     = {10.1007/s10278-013-9622-7}
}

@article{niyazi2023estro,
  title={ESTRO-EANO guideline on target delineation and radiotherapy details for glioblastoma},
  author={Niyazi, Maximilian and Andratschke, Nicolaus and Bendszus, Martin and Chalmers, Anthony J and Erridge, Sara C and Galldiks, Norbert and Lagerwaard, Frank J and Navarria, Pierina and af Rosensch{\"o}ld, Per Munck and Ricardi, Umberto and others},
  journal={Radiotherapy and Oncology},
  volume={184},
  pages={109663},
  year={2023},
  publisher={Elsevier}
}

@article{tran2025novel,
  title={Novel radiotherapy target definition using AI-driven predictions of glioblastoma recurrence from metabolic and diffusion MRI},
  author={Tran, Nate and Luks, Tracy L and Li, Yan and Jakary, Angela and Ellison, Jacob and Liu, Bo and Adegbite, Oluwaseun and Nair, Devika and Kakhandiki, Pranav and Molinaro, Annette M and others},
  journal={npj Digital Medicine},
  volume={8},
  number={1},
  pages={508},
  year={2025},
  publisher={Nature Publishing Group UK London}
}

@article{cepeda2023predicting,
  title={Predicting regions of local recurrence in glioblastomas using voxel-based radiomic features of multiparametric postoperative MRI},
  author={Cepeda, Santiago and Luppino, Luigi Tommaso and P{\'e}rez-N{\'u}{\~n}ez, Angel and Solheim, Ole and Garc{\'\i}a-Garc{\'\i}a, Sergio and Velasco-Casares, Mar{\'\i}a and Karlberg, Anna and Eikenes, Live and Sarabia, Rosario and Arrese, Ignacio and others},
  journal={Cancers},
  volume={15},
  number={6},
  pages={1894},
  year={2023},
  publisher={Mdpi}
}

@article{cepeda2023evaluation,
    author = {Cepeda, Santiago and Luppino, Luigi and Wodsinski, Marek and Solheim, Ole and Perez-Nuñez, Angel and Garcia-Garcia, Sergio and Karlberg, Anna and Eikenes, Live and Zamora, Tomas and Sarabia, Rosario and Arrese, Ignacio and Kuttner, Samuel},
    title = {NIMG-45. EXTERNAL EVALUATION OF A MACHINE LEARNING MODEL EMPLOYING RADIOMICS TO IDENTIFY REGIONS OF LOCAL RECURRENCE IN GLIOBLASTOMA FROM POSTOPERATIVE MRI},
    journal = {Neuro-Oncology},
    volume = {25},
    number = {Supplement 5},
    pages = {v195-v196},
    year = {2023}
}

@article{cepeda2023machine,
  title={Machine Learning-based Identification of Local Recurrence Regions in Glioblastoma using Postoperative MRI: Implications for Survival Prognostication},
  author={Cepeda, Santiago and Luppino, Luigi Tommaso and Solheim, Ole and P{\'e}rez-N{\'u}{\~n}ez, Angel and Garc{\'\i}a-Garc{\'\i}a, Sergio and Karlberg, Anna and Eikenes, Live and Zamora, Tomas and Sarabia, Rosario and Arrese, Ignacio and others},
  journal={Brain and Spine},
  volume={3},
  pages={101960},
  year={2023},
  publisher={Elsevier}
}

@article{isensee2021nnu,
  title={nnU-Net: a self-configuring method for deep learning-based biomedical image segmentation},
  author={Isensee, Fabian and Jaeger, Paul F and Kohl, Simon AA and Petersen, Jens and Maier-Hein, Klaus H},
  journal={Nature methods},
  volume={18},
  number={2},
  pages={203--211},
  year={2021},
  publisher={Nature Publishing Group}
}

@article{lasocki2019non,
  title={Non-contrast-enhancing tumor: a new frontier in glioblastoma research},
  author={Lasocki, A and Gaillard, F},
  journal={American Journal of Neuroradiology},
  volume={40},
  number={5},
  pages={758--765},
  year={2019},
  publisher={American Journal of Neuroradiology}
}

@article{chamberlain2011radiographic,
  title={Radiographic patterns of relapse in glioblastoma},
  author={Chamberlain, Marc C},
  journal={Journal of Neuro-Oncology},
  volume={101},
  number={2},
  pages={319--323},
  year={2011},
  publisher={Springer}
}

@article{soffietti2023neurotoxicity,
  title={Neurotoxicity from old and new radiation treatments for brain tumors},
  author={Soffietti, Riccardo and Pellerino, Alessia and Bruno, Francesco and Mauro, Alessandro and Rud{\`a}, Roberta},
  journal={International Journal of Molecular Sciences},
  volume={24},
  number={13},
  pages={10669},
  year={2023},
  publisher={MDPI}
}

@article{cepeda2025radiomics,
  title={Radiomics-based quantification of tumor infiltration in the non-enhancing peritumoral region on postoperative MRI is associated with survival in glioblastoma},
  author={Cepeda, Santiago and Esteban-Sinovas, Olga and Luppino, Luigi Tommaso and Kuttner, Samuel and Wodzinski, Marek and Solheim, Ole and Romero, Roberto and P{\'e}rez-N{\'u}{\~n}ez, Angel and Eikenes, Live and Karlberg, Anna and others},
  journal={Scientific Reports},
  volume={15},
  number={1},
  pages={43932},
  year={2025},
  publisher={Nature Publishing Group UK London}
}

@article{baheti2021brain,
  title={The brain tumor sequence registration (brats-reg) challenge: Establishing correspondence between pre-operative and follow-up mri scans of diffuse glioma patients},
  author={Baheti, Bhakti and Chakrabarty, Satrajit and Akbari, Hamed and Bilello, Michel and Wiestler, Benedikt and Schwarting, Julian and Calabrese, Evan and Rudie, Jeffrey and Abidi, Syed and Mousa, Mina and others},
  journal={arXiv preprint arXiv:2112.06979},
  year={2021}
}

@article{raissi2019physics,
  title={Physics-informed neural networks: A deep learning framework for solving forward and inverse problems involving nonlinear partial differential equations},
  author={Raissi, Maziar and Perdikaris, Paris and Karniadakis, George E},
  journal={Journal of Computational physics},
  volume={378},
  pages={686--707},
  year={2019},
  publisher={Elsevier}
}

@article{li2009solving,
  title={Solving PDEs in complex geometries: a diffuse domain approach},
  author={Li, Xiangrong and Lowengrub, John and R{\"a}tz, Andreas and Voigt, A25128341178},
  journal={Communications in mathematical sciences},
  volume={7},
  number={1},
  pages={81},
  year={2009}
}

@article{wen2020glioblastoma,
  title={Glioblastoma in adults: a Society for Neuro-Oncology (SNO) and European Society of Neuro-Oncology (EANO) consensus review on current management and future directions},
  author={Wen, Patrick Y and Weller, Michael and Lee, Eudocia Quant and Alexander, Brian M and Barnholtz-Sloan, Jill S and Barthel, Floris P and Batchelor, Tracy T and Bindra, Ranjit S and Chang, Susan M and Chiocca, E Antonio and others},
  journal={Neuro-oncology},
  volume={22},
  number={8},
  pages={1073--1113},
  year={2020},
  publisher={Oxford University Press US}
}

@inproceedings{mok2022unsupervised,
  title={Unsupervised deformable image registration with absent correspondences in pre-operative and post-recurrence brain tumor mri scans},
  author={Mok, Tony CW and Chung, Albert CS},
  booktitle={International Conference on Medical Image Computing and Computer-Assisted Intervention},
  pages={25--35},
  year={2022},
  organization={Springer}
}

@inproceedings{mok2022robust,
  title={Robust image registration with absent correspondences in pre-operative and follow-up brain MRI scans of diffuse glioma patients},
  author={Mok, Tony CW and Chung, Albert CS},
  booktitle={International MICCAI Brainlesion Workshop},
  pages={231--240},
  year={2022},
  organization={Springer}
}

@article{brooks2021white,
  title={The white matter is a pro-differentiative niche for glioblastoma},
  author={Brooks, Lucy J and Clements, Melanie P and Burden, Jemima J and Kocher, Daniela and Richards, Luca and Devesa, Sara Castro and Zakka, Leila and Woodberry, Megan and Ellis, Michael and Jaunmuktane, Zane and others},
  journal={Nature communications},
  volume={12},
  number={1},
  pages={2184},
  year={2021},
  publisher={Nature Publishing Group UK London}
}

@article{sansone2023patterns,
  title={Patterns of gray and white matter functional networks involvement in glioblastoma patients: indirect mapping from clinical MRI scans},
  author={Sansone, Giulio and Pini, Lorenzo and Salvalaggio, Alessandro and Gaiola, Matteo and Volpin, Francesco and Baro, Valentina and Padovan, Marta and Anglani, Mariagiulia and Facchini, Silvia and Chioffi, Franco and others},
  journal={Frontiers in Neurology},
  volume={14},
  pages={1175576},
  year={2023},
  publisher={Frontiers Media SA}
}

% ----------------------------------------------------------------------------------------------------
% Acknowledgements
% ----------------------------------------------------------------------------------------------------
\section{Acknowledgements}
Ray Zirui Zhang was supported in part by an NVIDIA academic award.

% ----------------------------------------------------------------------------------------------------
% Competing Interests
% ----------------------------------------------------------------------------------------------------
\section{Competing interests}
The GlioMap algorithm has been formally disclosed through a Disclosure of Invention (DOFI) filed with the University Hospital of Northern Norway (UNN), reference DOFI 2024/8833. The authors report no financial interests or royalties associated with this disclosure.

\clearpage
\section{Supplementary Information}
\label{sec:supplementary}

\subsection{Additional metrics}
\label{supp:metrics}
\begin{figure}[t]
\centering
\includegraphics[width=\linewidth]{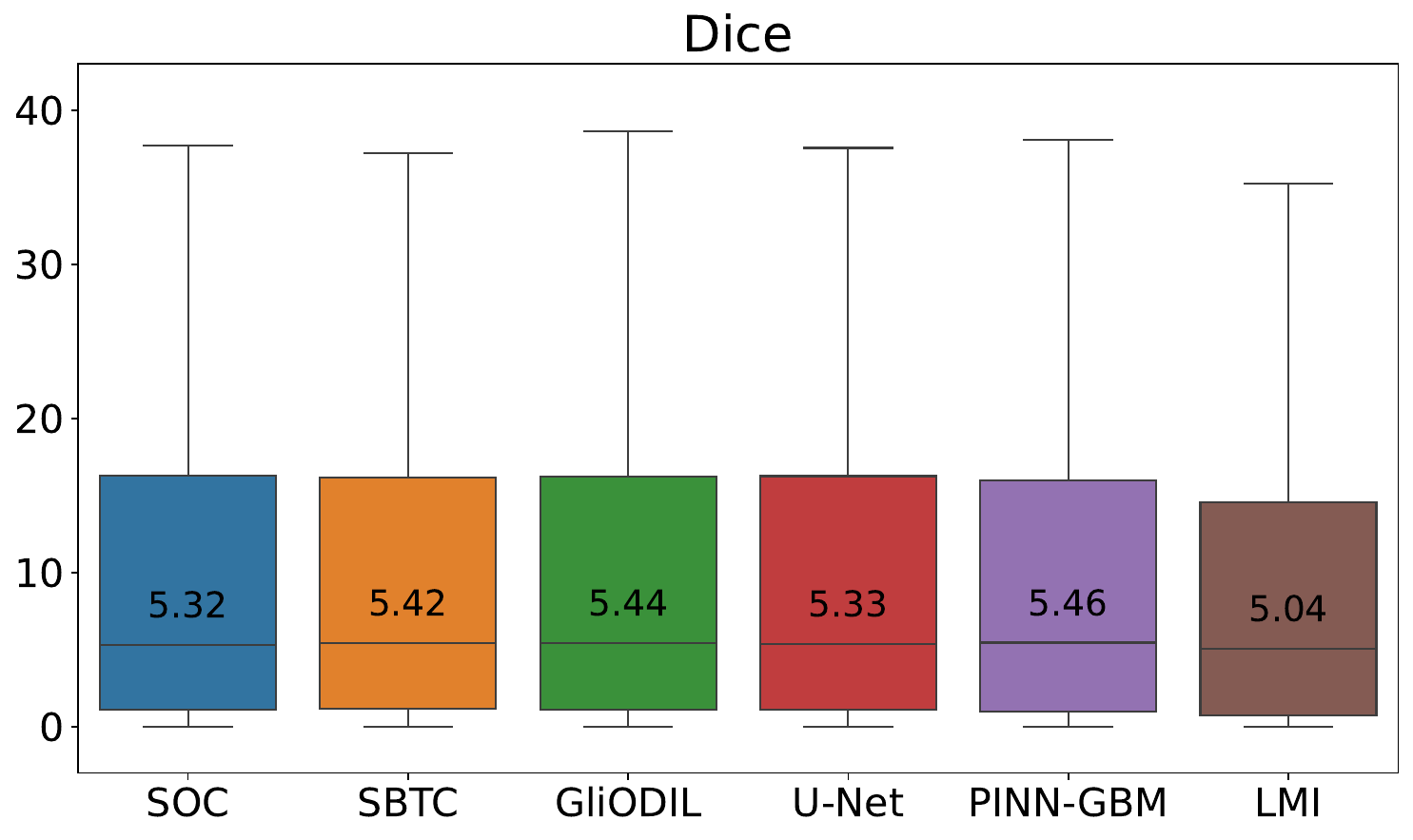}
\caption{Dice coefficient across the combined cohort for the enhancing recurrence.}
\label{fig:supp:dice}
\end{figure}

\begin{figure}[!t]
\includegraphics[width=\linewidth]{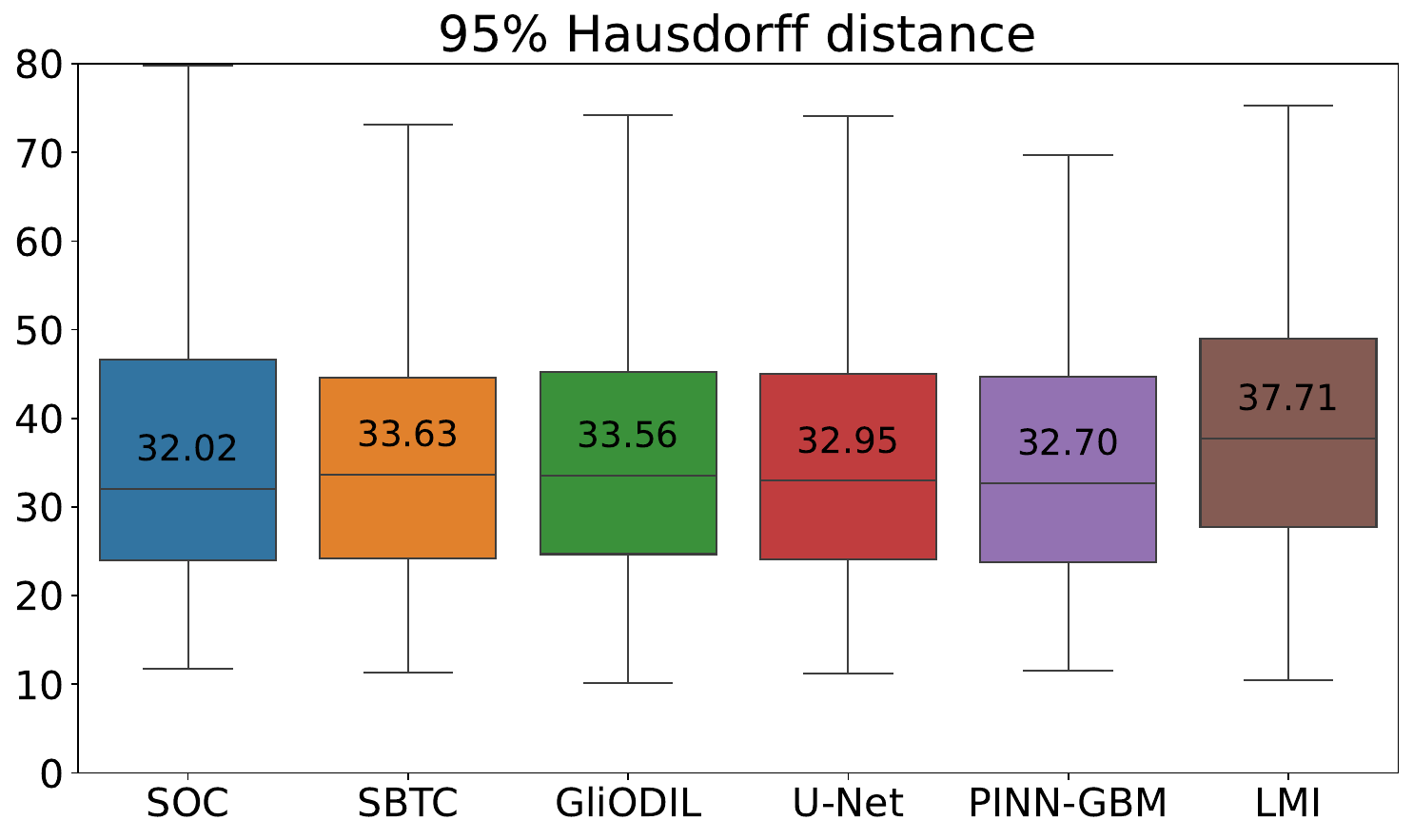}
\caption{95~\% Hausdorff distance across the combined cohort for the enhancing recurrence.}
\label{fig:supp:hausdorff}
\end{figure}

Figures~\ref{fig:supp:dice} and \ref{fig:supp:hausdorff} present the Dice coefficient and 95~\% Hausdorff distance, including median values, for the enhancing recurrence across the combined cohort. PINN-GBM, GliODIL, \ac{loti}, and our custom U-Net achieved slightly higher median Dice scores as well as slightly higher mean Dice scores. Our U-Net achieved the highest mean Dice score ($10.67 \pm 0.78$ compared to SOC $10.34 \pm 0.77$). For the 95~\% Hausdorff distance, \ac{soc} achieved the lowest (best) median value of $32.02\,\mathrm{mm}$. The best mean value was achieved by PINN-GBM with $36.44 \pm 1.21\,\mathrm{mm}$.

\subsection{Custom U-Net training details}
\label{supp:unet_details}
Our custom U-Net is trained using a deep-ensemble strategy with three ensemble members, each using different random initialization and randomized data shuffling~\cite{ronneberger2015u, lakshminarayanan_simple_2017}. For each ensemble member, we trained five U-Nets on distinct train/validation/test splits, resulting in a total of 15 trained U-Nets. To obtain unbiased out-of-sample predictions for all patients, each prediction is generated from the model for which the corresponding patient belongs to the held-out test set, ensuring that all results are obtained on unseen data. We used an extended dataset of 327 patients, yielding 197 training, 65 validation, and 65 test cases per split, stratified by dataset origin. The extended training dataset is supplemented with 84 additional patients from the \ac{cptac}~\cite{CPTAC_GBM_2018}, \ac{ivygap}~\cite{IvyGAP_2016}, \ac{tcgagbm}~\cite{TCGA_GBM_2016}, and \ac{upenngbm}~\cite{UPENN_GBM_2021}.

We use the following input imaging modalities: white matter, gray matter, and \ac{csf} probability maps, preoperative tumor segmentation map, preoperative \ac{t1c} and \ac{flair} scans, and \ac{adc} scans. \ac{adc} was available for 170 patients, with the remaining patients represented by zero-filled tensors. A binarized tumor recurrence segmentation, registered to the preoperative space, was used as ground-truth.

Our implementation uses base number of 32 features and a softmax activation to output tumor recurrence probabilities. Following~\cite{isensee2021nnu}, we use instance normalization~\cite{ulyanov_instance_2017} to prevent contrast shifting across batches with varying intensity ranges. Post-processing included Gaussian smoothing with $\sigma = 1$ to remove noise from the predictions. The U-Net was trained using a modified composite loss function inspired by~\cite{tran2025novel}, combining a binary cross entropy (BCE) loss with an individualized progression coverage coefficient (PCC) derived from the Tversky loss. The PCC loss dynamically adjusts loss hyperparameter values for each patient based on tumor size to address class imbalance. \ac{bce} and \ac{pcc} losses are weighted equally with a factor of $0.5$. We customized this composite loss by weighting recurring edema voxels with $0.25$ and recurring enhancing core voxels and necrosis voxels with $0.75$. Additionally, we apply a mask, setting the loss to zero for all voxels outside the brain.

We performed a grid search yielding a mini-batch size of 7, a learning rate $1 \times 10^{-4}$ with the Adam optimizer and weight decay $1 \times 10^{-4}$. To reduce the effect of overfitting, we employed a data augmentation framework largely inspired by~\cite{isensee2021nnu}. Augmentations were organized into image intensity augmentations, geometry augmentations, and label smoothing. We evaluated the impact of each category individually and found that all three contributed to improved recurrence coverage.

\textbf{Image Intensity Augmentations} were applied at random to either all input modalities or a randomly selected subset. These included downsampling, random intensity scaling, random Gaussian smoothing, random Gaussian noise, gamma augmentation, contrast transformation, and channel dropout.

\textbf{Geometry augmentations} were applied synchronously across input images and their pertinent brain mask and labels, including random flipping and random affine transformations.

\textbf{Label Smoothing Augmentations} were applied exclusively to labels to reduce label noise. We used random Gaussian edge softening~\cite{pornvoraphat2023real}, transforming the binary labels into continuous probability maps.

We furthermore regularized the predictions to avoid excessive deviation from the standard plan by generating a distance transformation of the center of mass of the tumor core and then linearly combining the distance transformation and the model output using weights $0.8$ and $0.2$, respectively. We found a small increase of $\sim 0.5~\%$ in performance from this step.

\end{document}